\documentclass[12pt,preprint]{aastex}

\usepackage{natbib}
\bibpunct{(}{)}{;}{a}{}{,}
\usepackage{lscape}

\slugcomment{}

\shorttitle{SINGS Serendipity observations of FIR cirrus emission}
\shortauthors{Bot C. et al.}

\begin{document}


\title{Serendipity observations of far infrared cirrus emission in the Spitzer Infrared Nearby Galaxies Survey:\\  Analysis of far-infrared correlations\altaffilmark{*}}

\altaffiltext{*}{This work is based on observations made with the \emph{Spitzer Space Telescope}, which is operated by the Jet Propulsion Laboratory, California Institute of Technology, under a contract with NASA.}


\author{Caroline Bot\altaffilmark{1,5} and George Helou\altaffilmark{1}  and Fran\c cois Boulanger\altaffilmark{2} and Guilaine Lagache\altaffilmark{2} and Marc-Antoine Miville-Deschenes\altaffilmark{2} and Bruce Draine\altaffilmark{3} and Peter Martin\altaffilmark{4}}
\email{bot@astro.u-strasbg.fr}


\altaffiltext{1}{California Institute of Technology, Pasadena CA 91125, USA}
\altaffiltext{2}{Institut d'Astrophyisque Spatiale, 91405 Orsay, FRANCE}
\altaffiltext{3}{Princeton University Observatory, Princeton, NJ08544, USA}
\altaffiltext{4}{Canadian Institute for Theoretical Astrophysics, Toronto, Ontario, M5S 3H8, Canada}
\altaffiltext{5}{Observatoire Astronomique de Strasbourg, 67000 Strasbourg, FRANCE}


\begin{abstract}
We present an analysis of far-infrared dust emission from diffuse cirrus clouds. This study is based on serendipitous observations at 160$\mu$m at high galactic latitude with the Multiband Imaging Photometer (MIPS) onboard the Spitzer Space Telescope by the Spitzer Infrared Nearby Galaxies Survey (SINGS). These observations are complemented with IRIS data at 100 and 60$\mu$m and constitute one of the most sensitive and unbiased samples of far infrared observations at small scale of diffuse interstellar clouds. 
Outside regions dominated by the cosmic infrared background fluctuations, we observe a substantial scatter in the 160/100 colors from cirrus emission. 
We compared the 160/100 color variations to 60/100 colors in the same fields and find a trend of decreasing 60/100 with increasing 160/100. This trend can not be accounted for by current dust models by changing solely the interstellar radiation field. It requires a significant change of dust properties such as grain size distribution or emissivity or a mixing of clouds in different physical conditions along the line of sight. These variations are important as a potential confusing foreground for extragalactic studies. \end{abstract}





\section{Introduction}

The InfraRed Astronomical Satellite (IRAS) showed for the first time that extended infrared emission was present at high galactic latitude, far from star forming regions \citep{Low:1984qy}. In these diffuse regions, clouds are optically thin to stellar radiation and the radiation field is relatively uniform which results in very limited variations of dust equilibrium temperature \citep{BAB+96,AOW+98,LABP98,Schlegel:1998ys}. These high latitude cirrus also show a tight correlation between their infrared emission (100$\mu$m to 1mm observed by DIRBE and FIRAS) and the HI column density \citep{BAB+96} and the dust emission is well characterized with a constant dust emissivity per hydrogen atom ($\tau/N_H=10^{-25}(\lambda/250)^{-2}cm^{2}$) close to the value expected from models of interstellar dust grains \citep{Draine:1984lr}. At shorter wavelength, the smaller dust grains emission is characterized by a ratio of $I_{60\mu m}/I_{100\mu m}\sim 0.2$ \citep{Laureijs:1991qy,Abergel:1996qy,BCJ00}. All in all, dust emission from local cirrus is then seen as rather homogeneous and simply characterized on large scales. However, little is known about the dust properties (e.g. optical properties for absorption and emission, distribution,\ldots) in these high latitude clouds, at resolutions smaller than the DIRBE beam (0.7$^o$).

Smaller scale analysis of infrared colors have been done on individual regions and show clear variations of dust properties. \citet{Laureijs:1996bs} and \citet{Abergel:1994fy} observed a decrease of $I_{60\mu m}/I_{100\mu m}$ toward dense clouds. \citet{Bernard:1999qy} studied the far infrared emission at the arcminute scale in the Polaris flare with IRAS, ISOPHOT and PRONAOS (200 to 600$\mu$m) in a region where extended emission from cirrus is detected as well as a denser structure. The spectrum of the extended cirrus indicates a low dust temperature associated with a low 60/100 $\mu$m ratio. This was  also observed in the Polaris flare toward moderately dense regions ($A_V\sim 1$) and in a denser filament in the Taurus complex \citep{CBL+01,SAB+03}. It might be explained by the formation of large dust aggregates through the adhesion of small dust particles onto the surface of larger grains, leading to a change of dust emissivity properties. In the dense regions the very small grains seem to have disappeared almost completely. However, all these observations were restricted to individual regions, most of which are much denser than the diffuse local interstellar medium seen at high galactic latitudes. 

By comparing near-infrared extinction and extinction deduced from far-infrared dust emission in the whole anticenter hemisphere, \citet{Cambresy:2005vl} observed a discrepancy between the two quantities in regions above 1 mag. This effect is also interpreted by a change of dust emissivity  due to the presence of fluffy grains and the grain-grain coagulation scenario was therefore extended to larger regions. 

\citet{Kiss:2006uq} analyzed the far-infrared emission properties in a large sample of interstellar clouds observed with ISOPHOT with respect to extinction in regions of the order of 100 arcmin$^2$. They find variations of the far infrared dust emissivities in the coldest (12K$<T_d<$14K) and densest regions that are consistent with a dust grain growth scenario. But they also observe changes of the dust emissivities in the warmer regions (14K$<T_d<$17.5K) and interpret them as an effect of mixing along the line of sight of components with different temperatures or a change of the dust grain size distribution. However, a fraction of their sample was chosen on the basis of high brightnesses in the IRAS bands and could therefore be biased toward regions with enhanced small grain emission. 

Extinction measures toward high galactic latitude sightlines show a substantial fraction of low $R_V=A_V/E(B-V)$ values, also indicative of enhanced relative abundances of small grains. However, with the lack of longer wavelength measurements at small scales, it is difficult to relate these variations to possible changes in the dust grain properties.

The photometric data from the Spitzer Space Telescope enable us to have access to sensitive observations up to 160$\mu$m. Among the large programs, the Spitzer Infrared Nearby Galaxies Survey (SINGS) observed a sample of 75 nearby galaxies in photometry with the IRAC and MIPS instruments. The fields observed were chosen to be at high galactic latitude in order to limit the foreground cirrus contamination in the study of the targeted galaxy. Because the region observed was larger than the targeted galaxy, these serendipitous observations are then ideal to study far infrared dust emission in a large sample of high galactic latitude regions. We combine these new observations with IRIS data \citep{ML05} at 60 and 100$\mu$m,  a reprocessing of IRAS data including a better calibration of the infrared brightnesses and better zodiacal light subtraction. The goal of this paper is to \emph{study the infrared colors of diffuse local dust emission on the scale of a few arcminutes.}
 
\section{The data}

\begin{figure*}
\plottwo{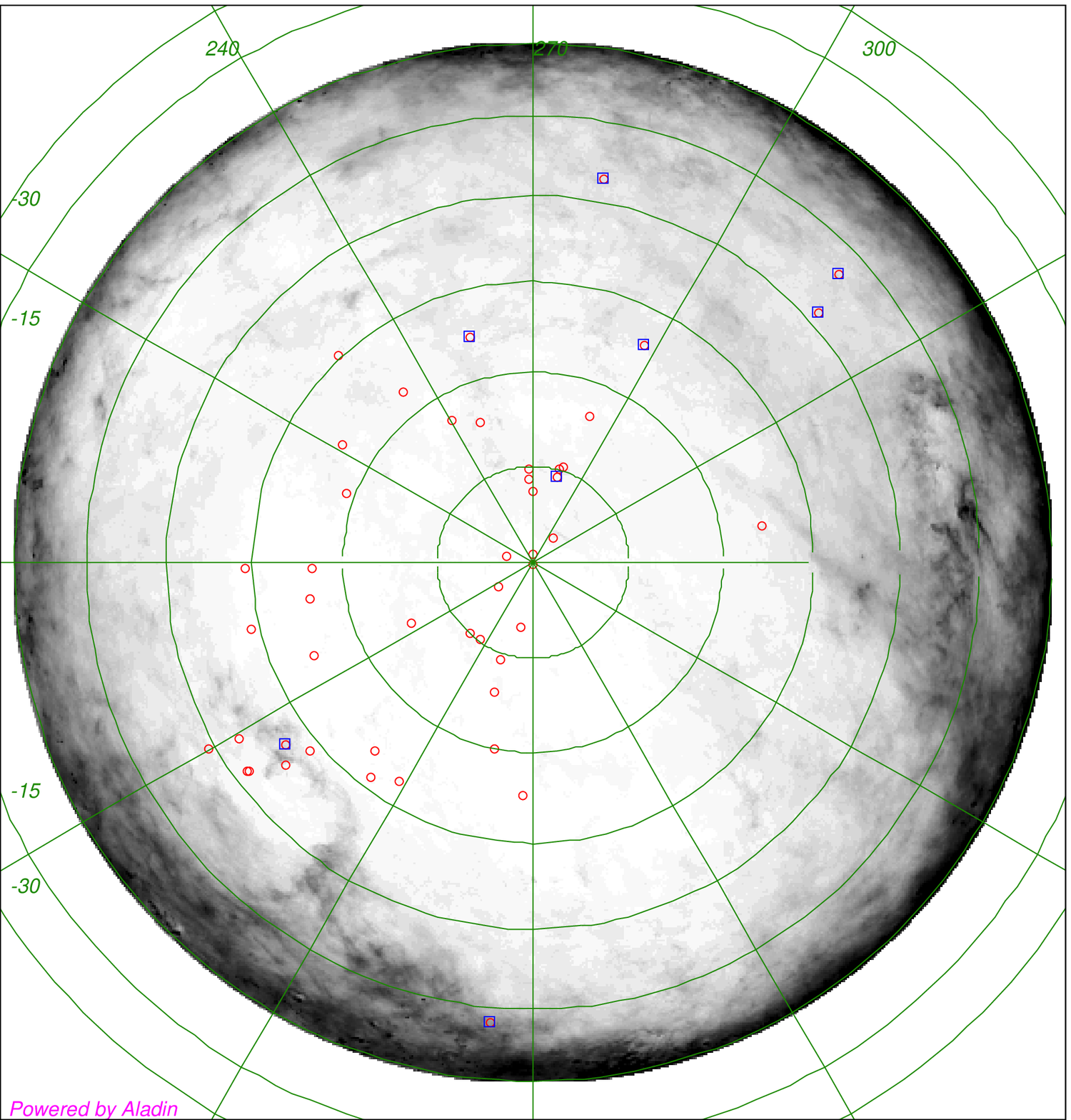}{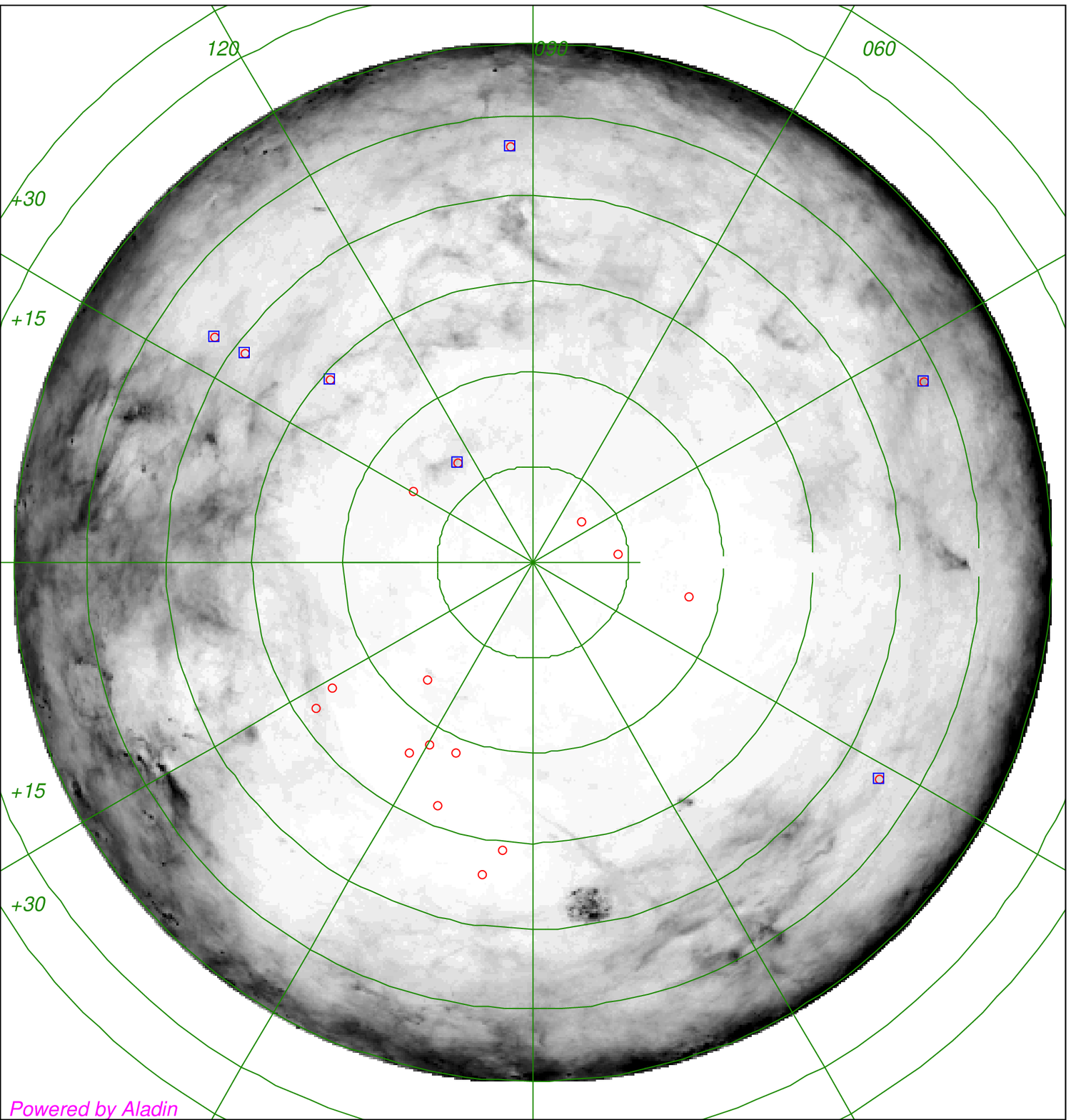}
\caption{Position of the SINGS fields (red circles and blue squares) overlaid on the dust column density maps from \citet{Schlegel:1998ys} centered around the north galactic pole (left panel) and the  south galactic pole (right panel). The blue squares correspond to the fields selected for this study (the variation in the infrared emission is dominated by the cirrus component).  A grid of galactic coordinates is overlayed. \label{fig0}}
\end{figure*}

The Spitzer Infrared Nearby Galaxies Survey \citep[SINGS][]{Kennicutt:2003fk} observed in imagery with IRAC \citep{Fazio:2004fj} and MIPS \citep{Rieke:2004zr} onboard Spitzer a sample of 75 nearby galaxies. While the IRAC images only observed the galaxy itself, a significant part of the MIPS observations (strips) encompass the surrounding sky. Since the SINGS observations were chosen to be at high galactic latitude to limit the galactic foreground contamination, the MIPS observations at 160$\mu$m provide a good opportunity to study the low surface brightness diffuse infrared emission from high galactic cirrus in a large number of fields at a resolution of $\sim 37"$. These data are complemented with IRIS data \citep{ML05} at 100 and 60$\mu$m. The position of the fields on the sky are shown in Fig. \ref{fig0} while their characteristics are summarized in Tab. \ref{tab1}. 

Although SINGS observations were also done at 70$\mu$m with MIPS, the regions observed are offset with respect to the galaxy targeted and only a small fraction of the 70 and 160$\mu$m observations overlap outside the galaxy itself, making them inappropriate for our galactic cirrus emission study. $24\mu$m observations were also available, but they are dominated by point sources emission as well as stronger zodiacal light. Once the point sources removed and the regions at low ecliptic latitude are discarded, the 24$\mu$m brightness have a low dynamic range in each field and no meaningful correlation can be done with longer wavelength observations. This study was therefore restricted to the comparison of 60, 100 and 160$\mu$m brightnesses. The IRIS 25$\mu$m observations were however used together with longer wavelength in order to remove point sources (like galaxies) more efficiently in the observations. 

\subsection{Data treatment \label{sec:dattreat}}

The observations at 160$\mu$m were reduced using the GeRT software\footnote{\url{http://ssc.spitzer.caltech.edu/mips/gert/index.html}} on the raw MIPS observations. Standard parameters were used for the reduction, but data where flashes of the internal source led to a significant number of saturated pixels which were removed. The removed data are most often positioned on the bright center of the galaxy. Other saturated pixels removed from processing were due to cosmic ray hits. These saturated flashes when not removed can bias the sensitivity of the diffuse extended emission. This step in the reduction may be not appropriate for the photometry of the galaxy, but significantly reduces latents (stripes) in the outer regions we are interested in. Each region targeted was observed twice. Discrepant fluxes at the same position between the two observations are removed and the data are combined into a mosaic for each region.

The MIPS 160$\mu$m maps and IRIS 60$\mu$m are convolved to the IRIS 100$\mu$m resolution assuming gaussian beams with FWHM of 4.0', 4.3' and 37" for IRIS 60, 100$\mu$m and MIPS 160$\mu$m observations respectively. 
 
For each MIPS strip, the galaxy and other point sources are detected in the 25, 60 and 100$\mu$m maps using the method described in \citet{Miville-Deschenes:2002vn}. These point sources at the IRIS resolution (but with the MIPS sampling) are then smoothed by a gaussian kernel with a full width half maximum of $3\times 3$ pixels (at the MIPS original pixel size) to encompass possible extended emission from these galaxies. The smoothed point sources are then masked in all the maps.  All maps are then projected on the IRIS grid to avoid oversampling. 
The observation targeting the galaxy Holmberg IX was removed from the sample since the emission in the whole strip is dominated by the galaxy and its interaction features with nearby galaxies. We chose to remove the observation containing the galaxy NGC3034 (M82) which was hampered by saturation effects in the whole central region of the galaxy, affecting the observation globally. The observations containing the galaxies NGC1266, NGC2915 and M81 Dwarf B were also removed because the width of the region observed were too narrow to be convolved meaningfully to the IRIS resolution. We ended up with 70 fields of view\footnote{Although there are 75 galaxies in the SINGS sample, some galaxies are in the same field of view: NGC3031 is in the same observation as M81 dwarf B and NGC5195 was observed simultaneously with NGC5194} at a resolution of 4.3' observed at 60, 100 and 160$\mu$m. Due to uncertainties in the zodiacal light subtraction at 60$\mu$m that can dominate the flux at the low surface brightnesses we sample, we limited the sample at 60$\mu$m to the 9 observations at high ecliptic latitude ($|\beta|>15^o$).

 A constant brightness of 0.78 MJy/sr is removed from the IRIS 100$\mu$m maps to account for the cosmic infrared background \citep{Lagache:2000lr}, i.e. the emission from the distant unresolved galaxies (called hereafter CIB). The exact level of CIB emission has not yet been established at 60$\mu$m and the MIPS observations can have offsets in the calibration of the brightness that are not physical. To overcome the uncertainties (physical or instrumental) on the zero levels in the different maps, we hereafter perform the analysis of the data through the use of correlations (c.f. \S \ref{sec:160and100compar}).

The errors on the surface brightness are taken to be 0.03 and 0.06 MJy/sr at 60 and 100$\mu$m respectively \citep{ML05}. At 160$\mu$m, we take a quadratic combination of a constant sensitivity limit of 0.12 MJy/sr \footnote{the sensitivity of the observations is computed for a 16s integration time per pixel using the SENS-PET tool, http://ssc.spitzer.caltech.edu/tools/senspet/ and is  divided by $\sqrt{N}$ where $N=49$ is the number of MIPS 160$\mu$m PSF inside an IRIS PSF at 100$\mu$m} and a 2\% uncertainty on the brightness (due to the uncertainty on the calibration factor from instrumental units to surface brightness \citep{Stansberry:2007fr}). 

\subsection{Sample selection}

In low surface brightness regions, the variations of the infrared emission in the observations can come from cirrus emission, fluctuations in the cosmic infrared background or from noise. Since we want to study the variations of cirrus emission only, we want to select observations in the SINGS sample that are dominated by the dust emission variations. 

\begin{figure}
\plotone{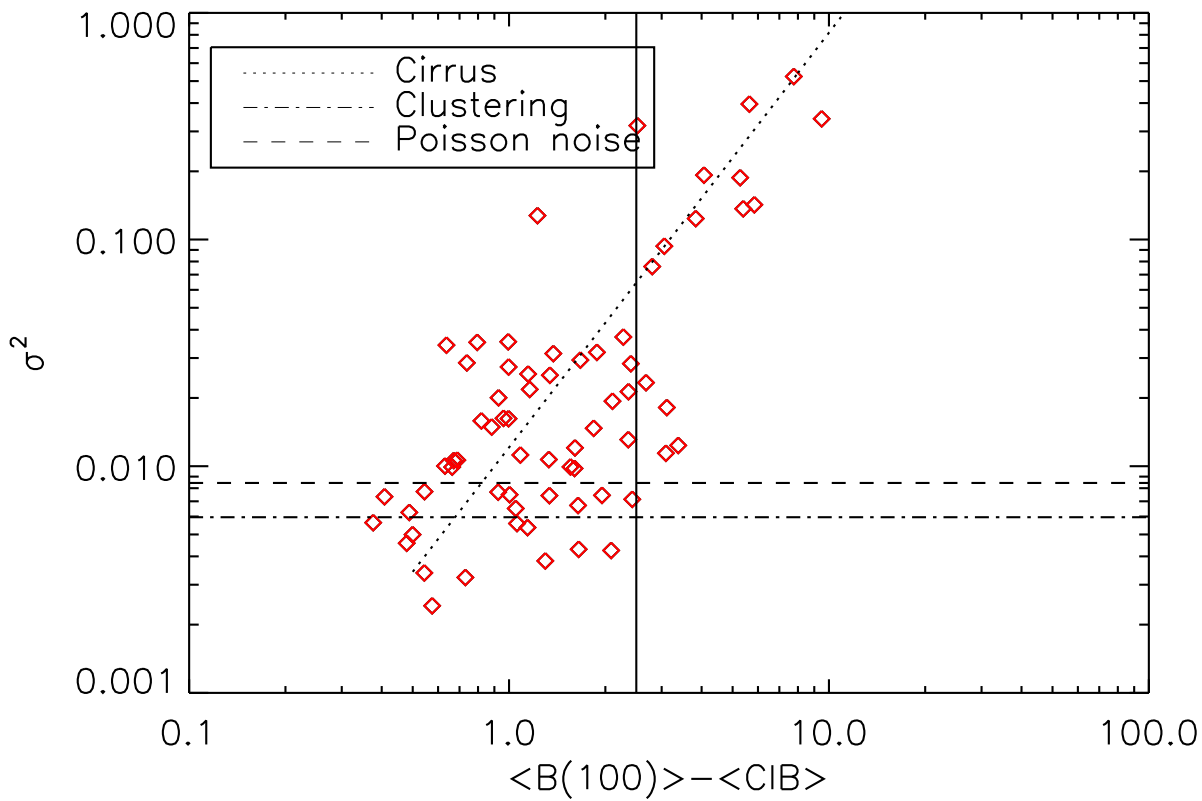}
\caption{Variations of the square of the standard deviation (related to the power spectrum of the signal) measured in each field with the mean brightness at 100$\mu$m. The observed values are compared with models for the different contributions: the infrared galaxies clustering (dotted-dashed line), the poisson noise (dashed line) and the cirrus variation (dotted line). This enables us to define a cut in 100$\mu$m brightness (the vertical black line) above which the cirrus variations dominate over CIB fluctuations. \label{fig1}}
\end{figure}

The different contributions (cosmic infrared background, cirrus emission) to the infrared emission have different power spectra that can help to disentangle them. In particular, the cirrus power spectrum normalization depends on the mean surface brightness while the contribution from background galaxies does not.  This dependence can be translated into a relationship between the mean brightness and the standard deviation square, $\sigma_{cirrus}^2$, in a region and depends on the size of the region \citep{Miville-Deschenes:2007yq}. For each of the observed field of view, we computed the standard deviation at 100$\mu$m and the mean brightness at 100$\mu$m (minus the average CIB contribution at this wavelength) and then plot the $\sigma^2$--$<B>$ relationship observed at 100$\mu$m in our sample. To model $\sigma_{cirrus}$, we use the relationship derived by \citet{Miville-Deschenes:2007yq} below 10 MJy/sr, for a maximum scale length of 50' (dotted line in Fig \ref{fig1}). We observe that our observations are consistent with the model, with a large scatter as in the original relationship.  This dispersion is likely enhanced due to the fact that our fields of view are elongated and the size of the region mapped varies between field.The contribution from the CIB fluctuations can be described by two terms: a Poisson noise that represents the galaxies distributed homogeneously with respect to the resolution and a component with correlated spatial variations corresponding to the clustering of galaxies on large scales. The contribution from the clustering of infrared galaxies is predicted by using the \citet{Lagache:2003rz} model for galaxy evolution, with a bias parameter from \citet{Lagache:2007qv}. The contribution from the Poisson noise to the $\sigma^2$ observed at 100$\mu$m is taken to be that measured by \citet{Miville-Deschenes:2002vn} since we used the same point source detection method. However, compared to their study, we removed point sources applying the detection scheme at all wavelength (25, 60 and 100$\mu$m). This enables us to mask faint galaxies at 100$\mu$m more efficiently and the Poisson noise in our measurements could be lower than their measurement. Because we want to select the fields with the least contribution from other sources (CIB) than cirrus to the observed variations, this choice is therefore conservative. 

Combining all contributions (represented in Fig. \ref{fig1}) to the observed variations, we determine that a cut at 2.5 MJy/sr corresponds to $\sigma_{cirrus}/\sigma_{CIB}=1.5$ (so that the total infrared fluctuations $\sigma_{tot}=\sqrt{\sigma_{cirrus}^2+\sigma_{CIB}^2}$ are less thatn 20\% larger than from cirrus fluctuations alone). The regions with a mean 100$\mu$m brightness above this threshold will therefore be dominated by variations of cirrus emission. In each field, we computed the mean brightness at 100$\mu$m as well as the standard deviation at 60, 100 and 160$\mu$m (c.f. Tab.\ref{tab1}).  By keeping only the fields above the 2.5 MJy/sr cut, the sub-sample we will study in this paper is composed of 15 fields with 100 and 160$\mu$m brightnesses, among which 9 can be studied as well at 60$\mu$m ($|\beta|>15^o$, see sec. \ref{sec:dattreat}).

Stellar reddenings obtained from the analysis of the Sloan Digital Sky Survey data enable us to put an upper limit of 1.2 mag on the extinction in these fields\footnote{Using $N(H)/A_V$ \citep{BSD78} and $B_{100}/N(H_I)\approx 6.67\times 10^{-21} MJy/sr cm^2$, this upper limit implies $B_{100}<15.2MJy/sr$, which is fully consistent with the brightness observed in our sample}. This confirms that the variations in the infrared cirrus emission studied in each region comes from diffuse clouds according to the \citet{van-Dishoeck:1988wd} classification.

\section{Results}

\subsection{Cirrus emission at 160 and 100$\mu$m\label{sec:160and100compar}}

\begin{figure*}
\includegraphics[scale=.30]{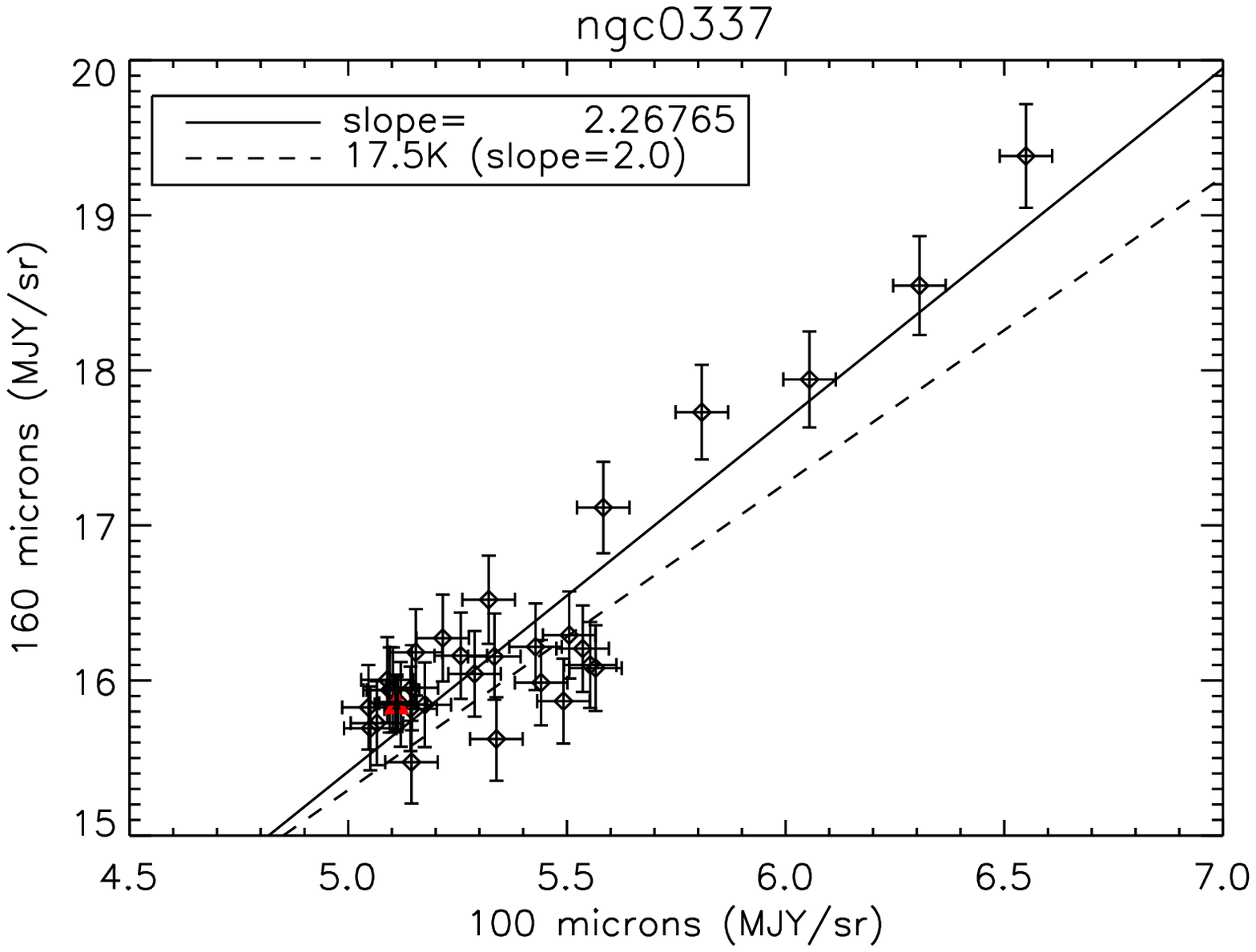}
\includegraphics[scale=.30]{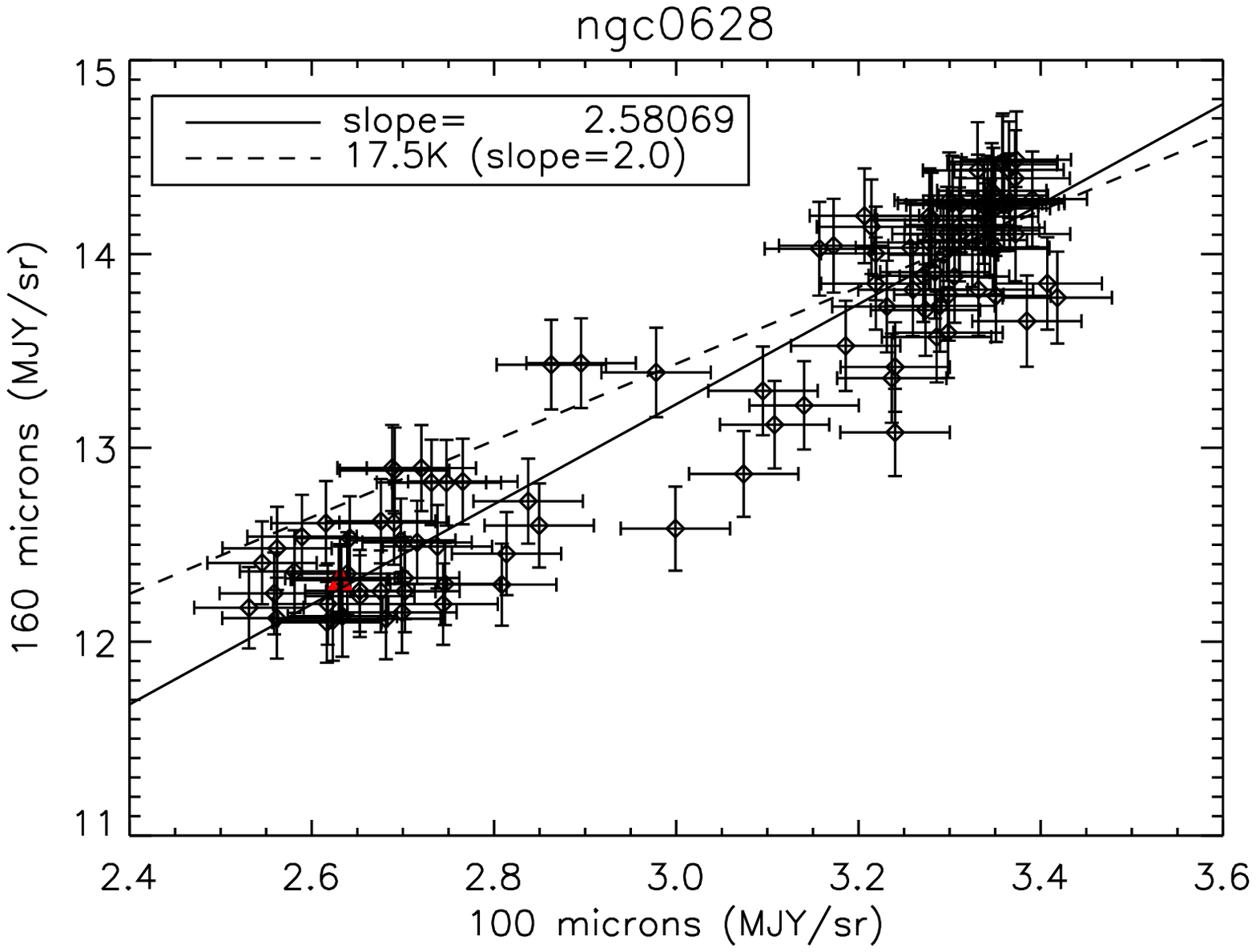}
\includegraphics[scale=.30]{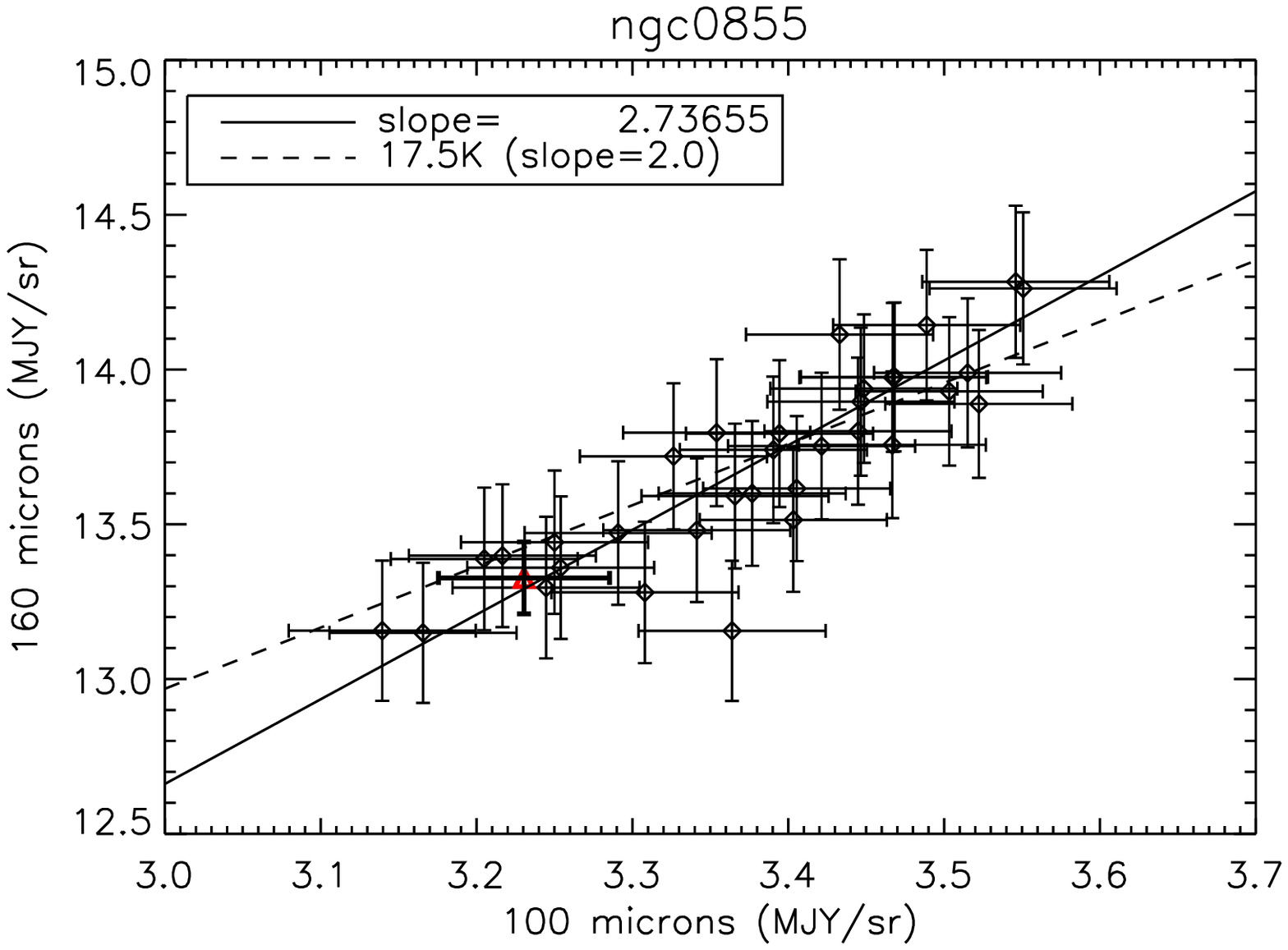}\\
\includegraphics[scale=.30]{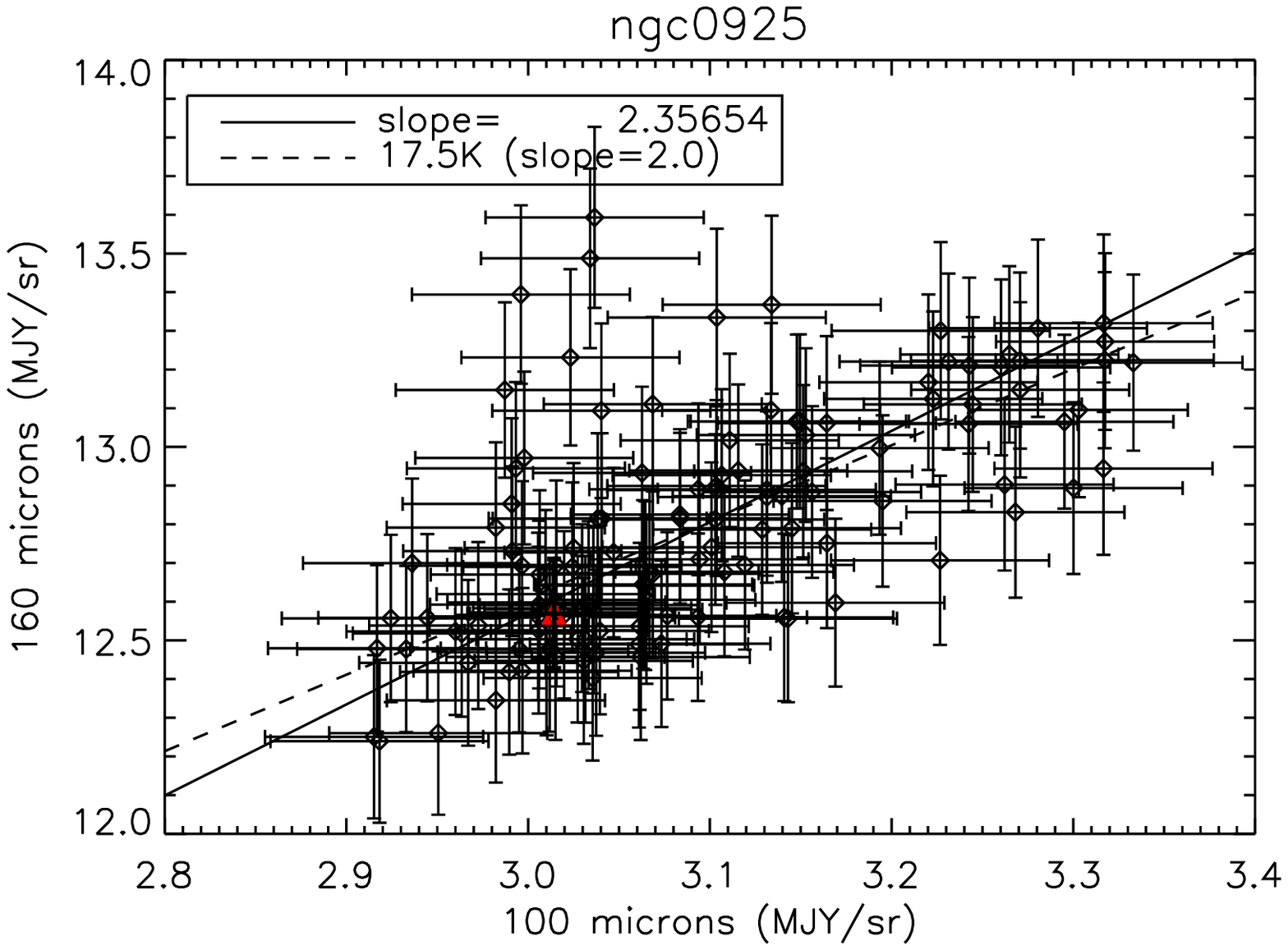}
\includegraphics[scale=.30]{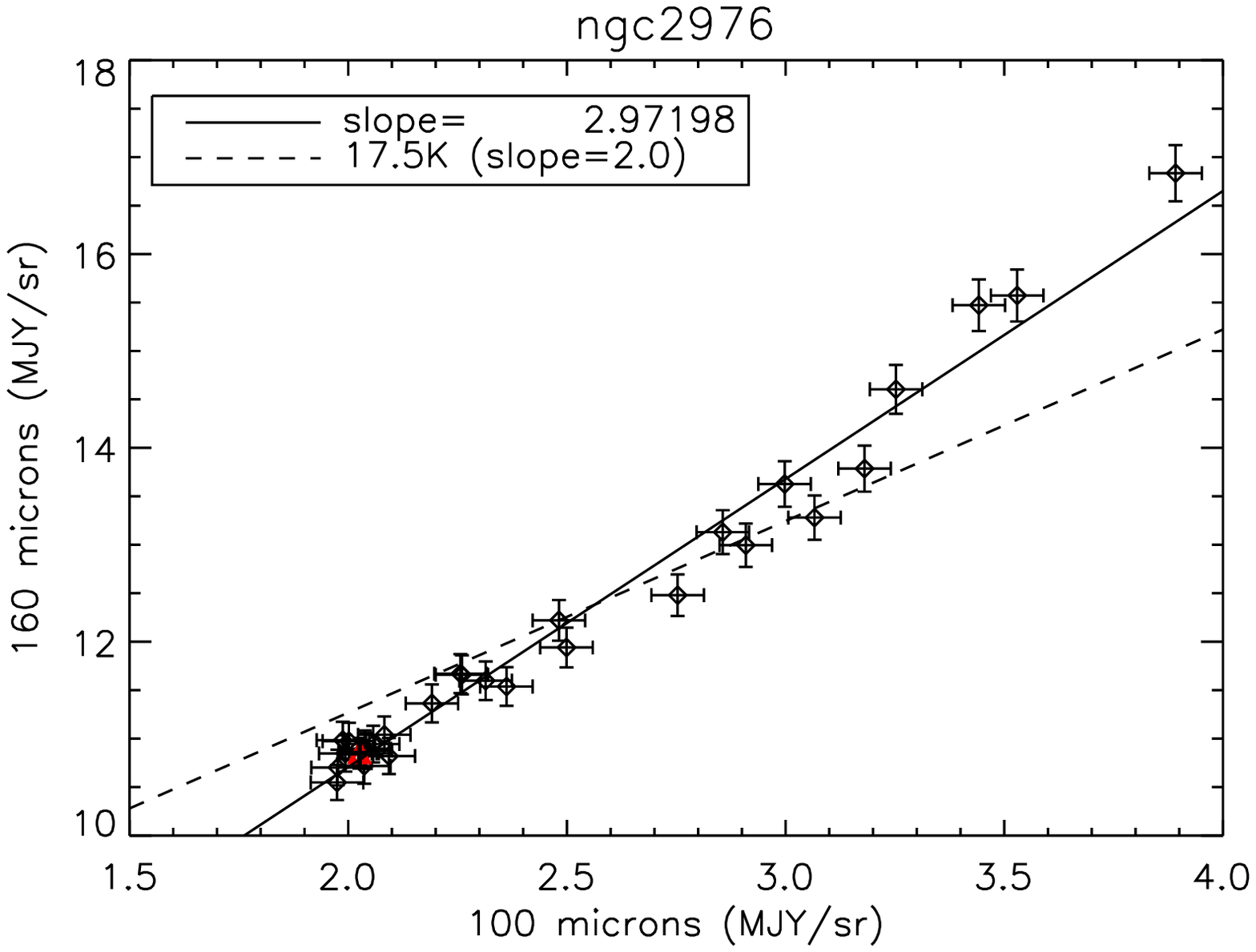}
\includegraphics[scale=.30]{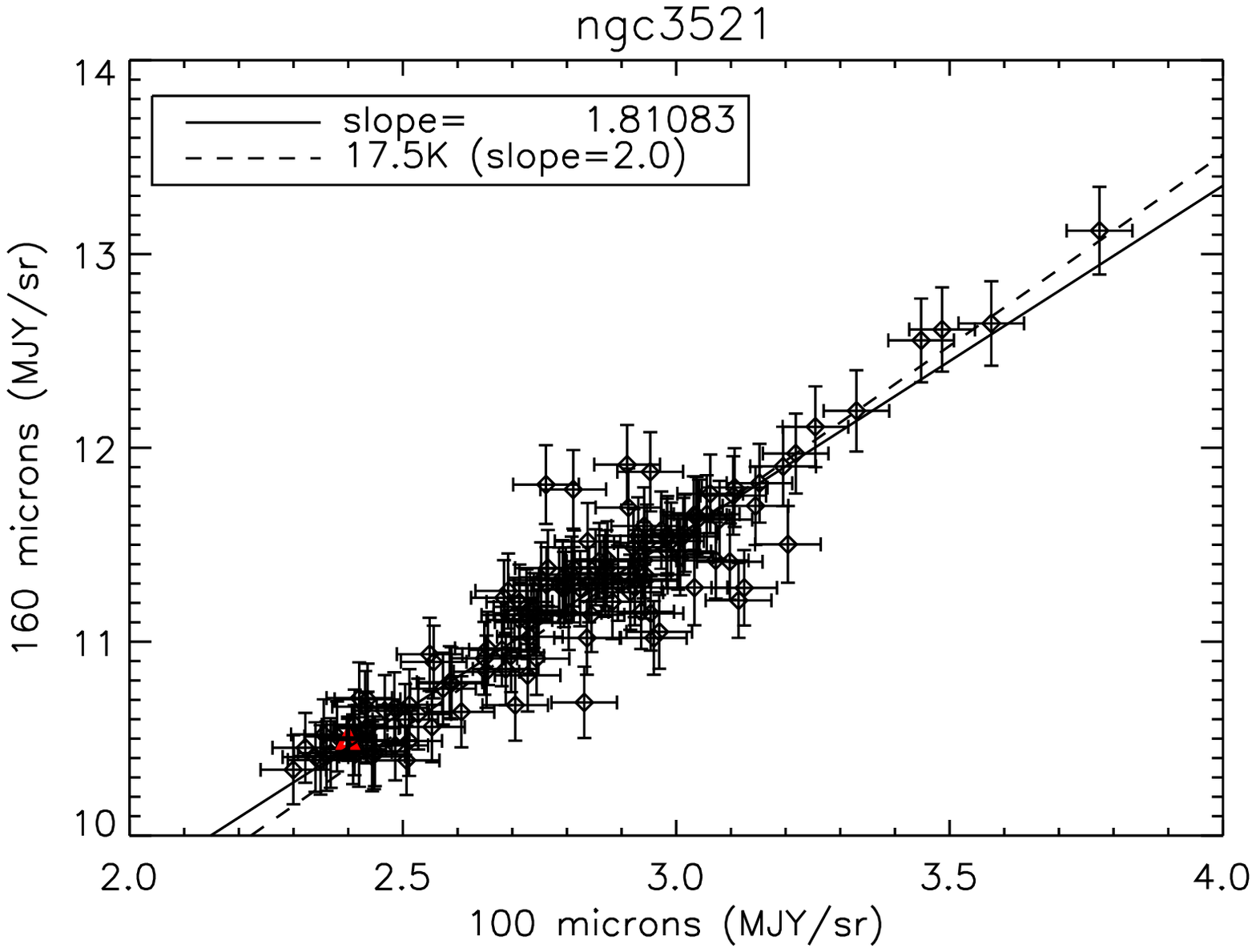}\\
\includegraphics[scale=.30]{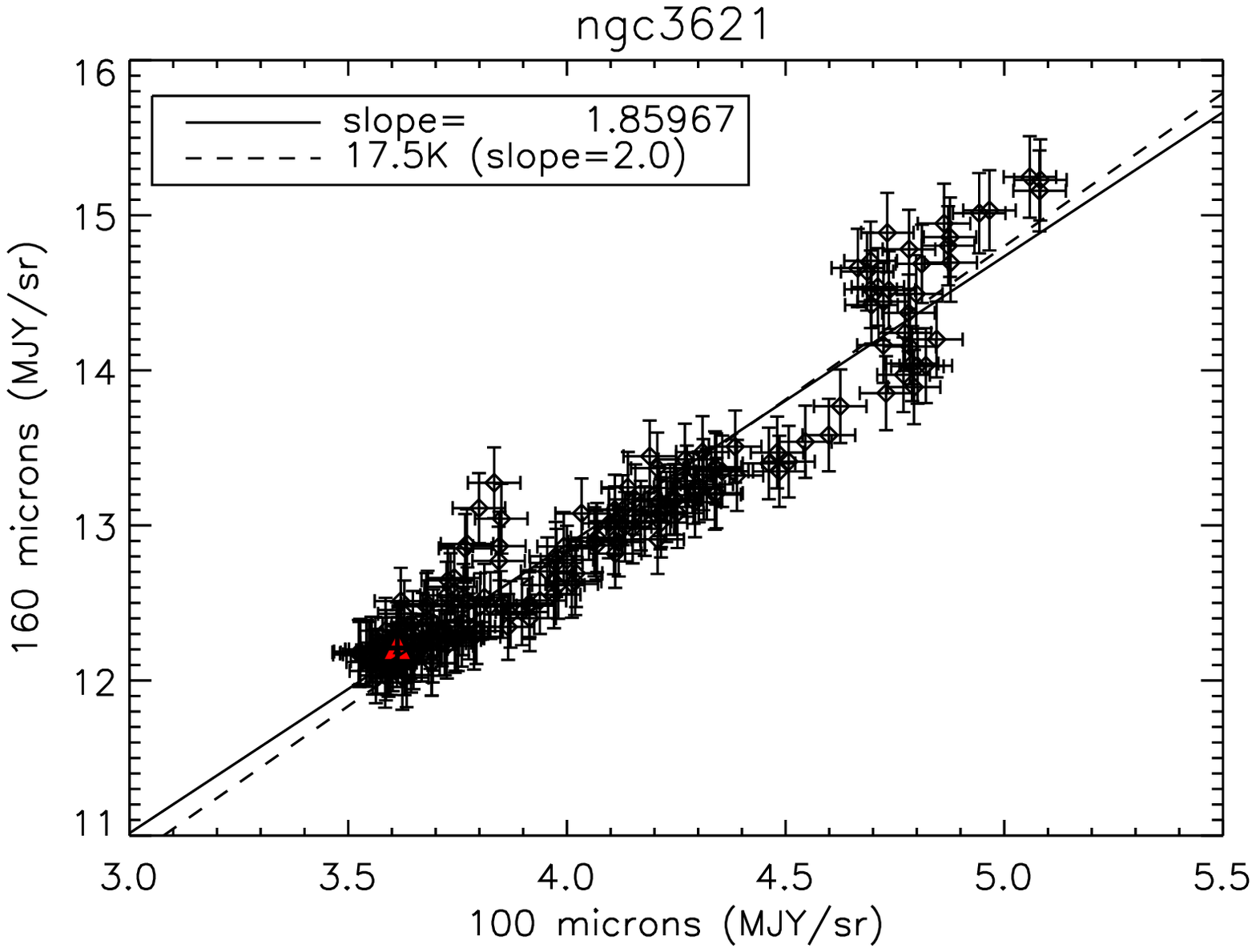}
\includegraphics[scale=.30]{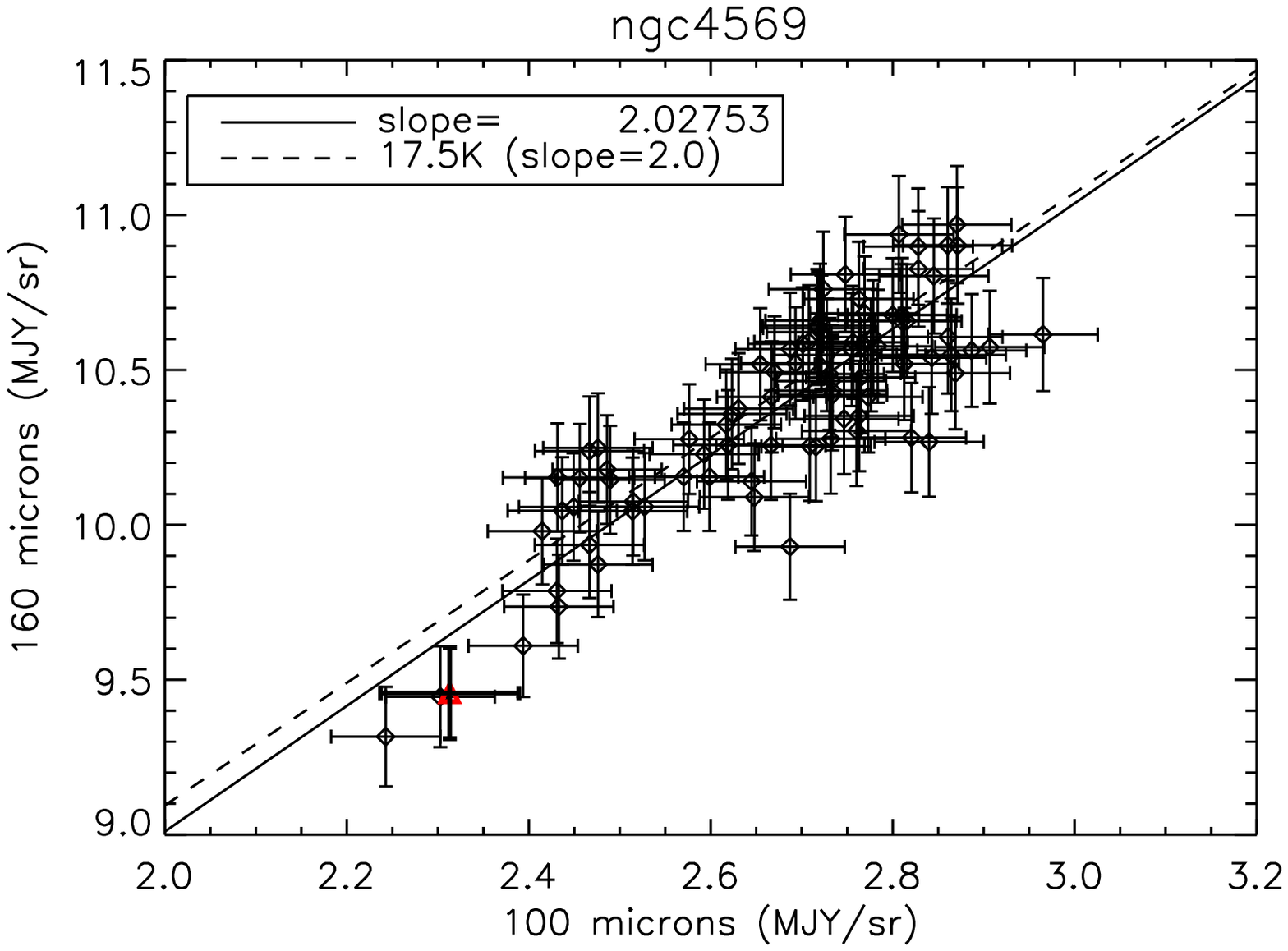}
\includegraphics[scale=.30]{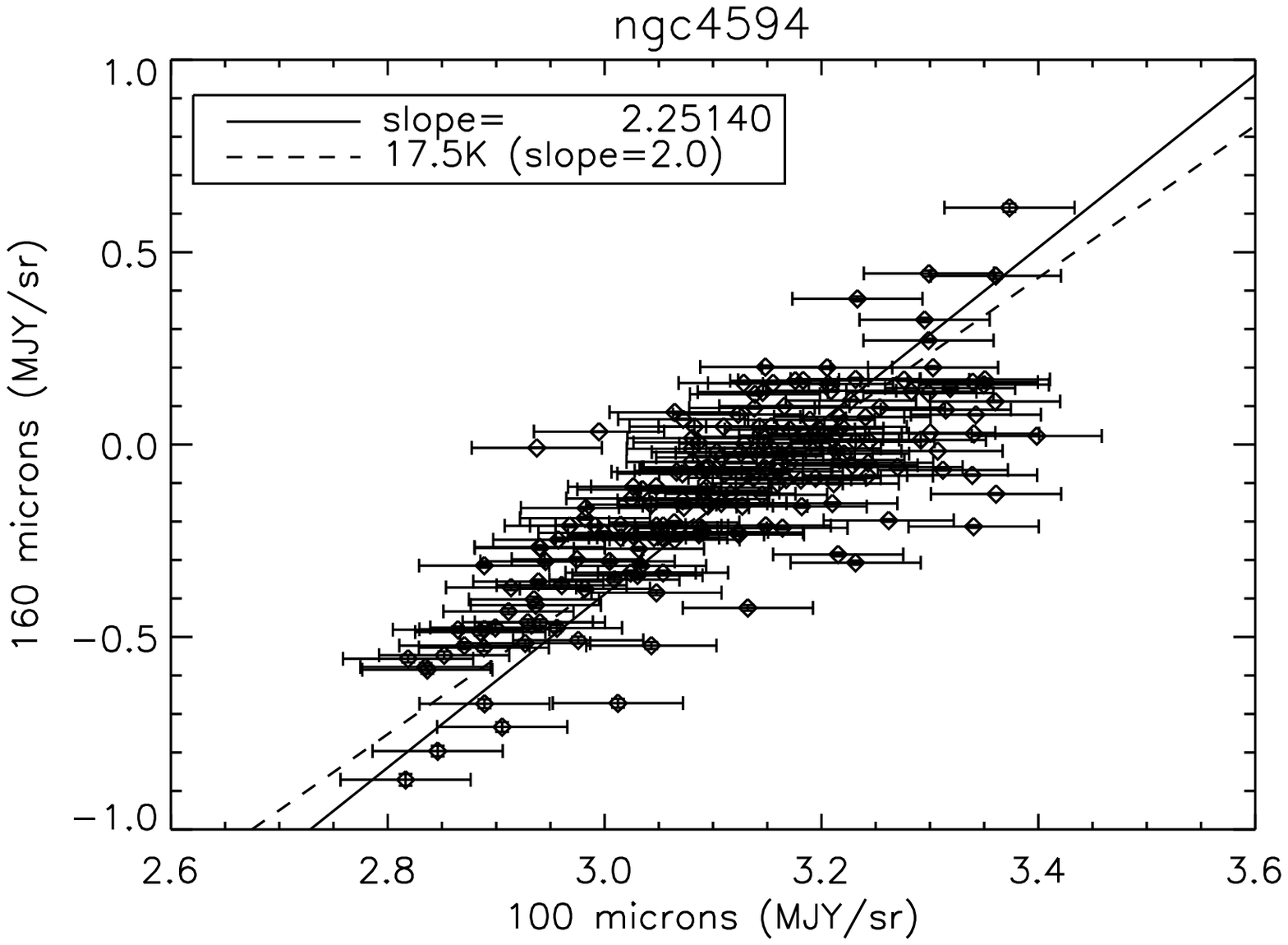}\\
\caption{ 160--100 scatterplots for all SINGS observations with $<B_{100}> \geq 2.45$ MJy/sr. In each plot, a canonical slope of 2.0 is represented by a dashed line (corresponding to a temperature of 17.5K). A linear fit is performed on the correlation and the best fit is represented with a solid line. The value of the slope obtained is written in the legend. \label{fig2}}
\end{figure*}

\begin{figure*}
\includegraphics[scale=.30]{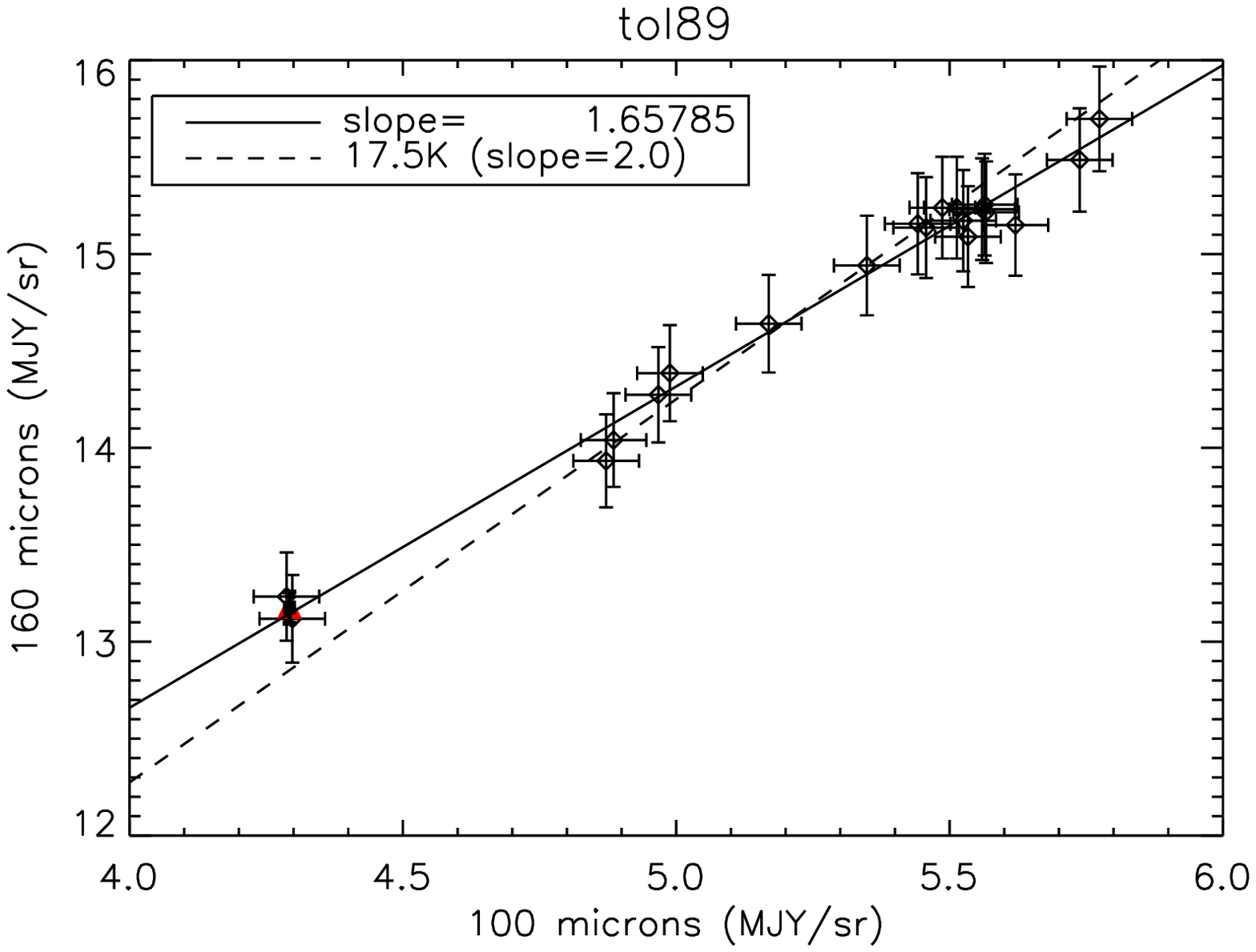}
\includegraphics[scale=.30]{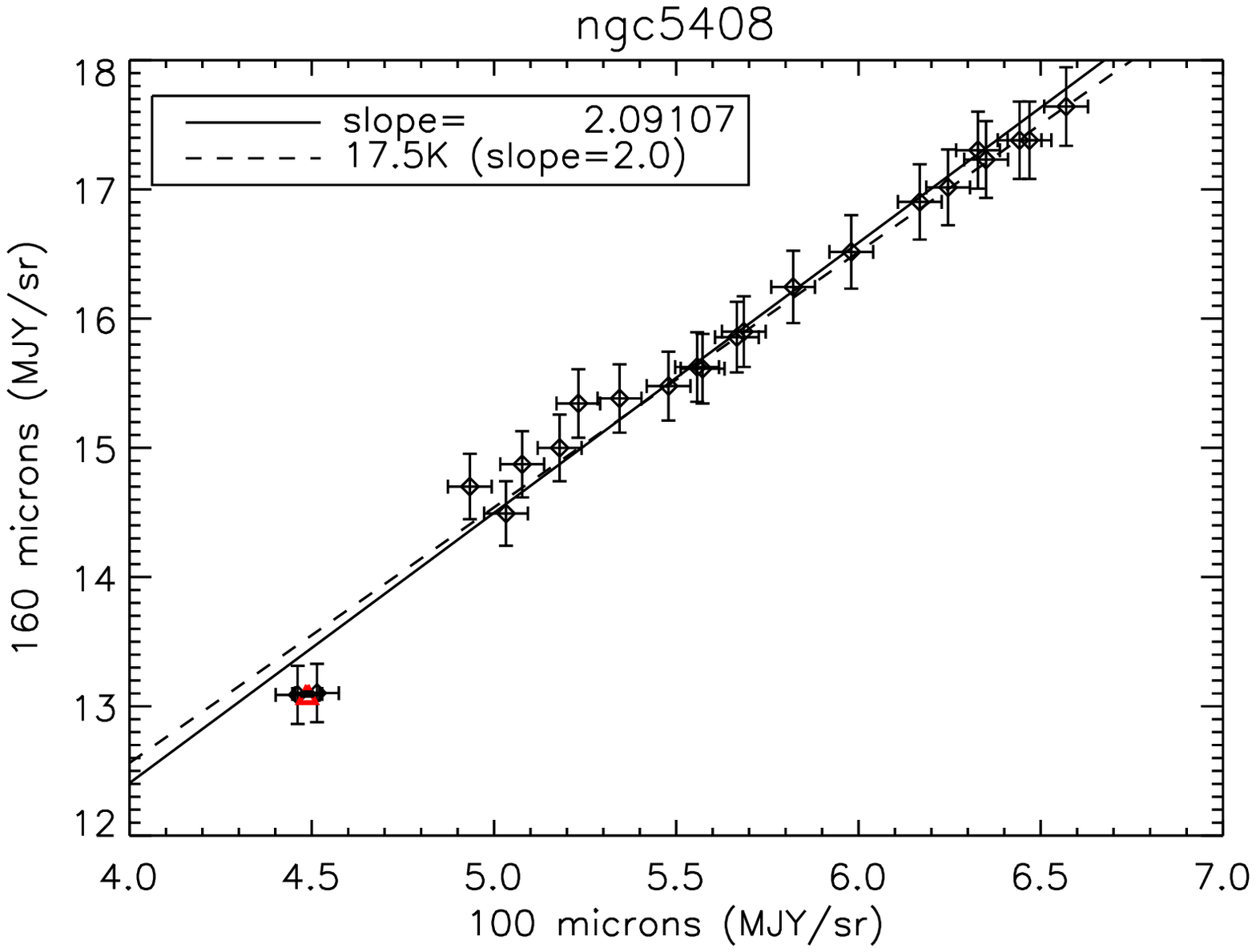}
\includegraphics[scale=.30]{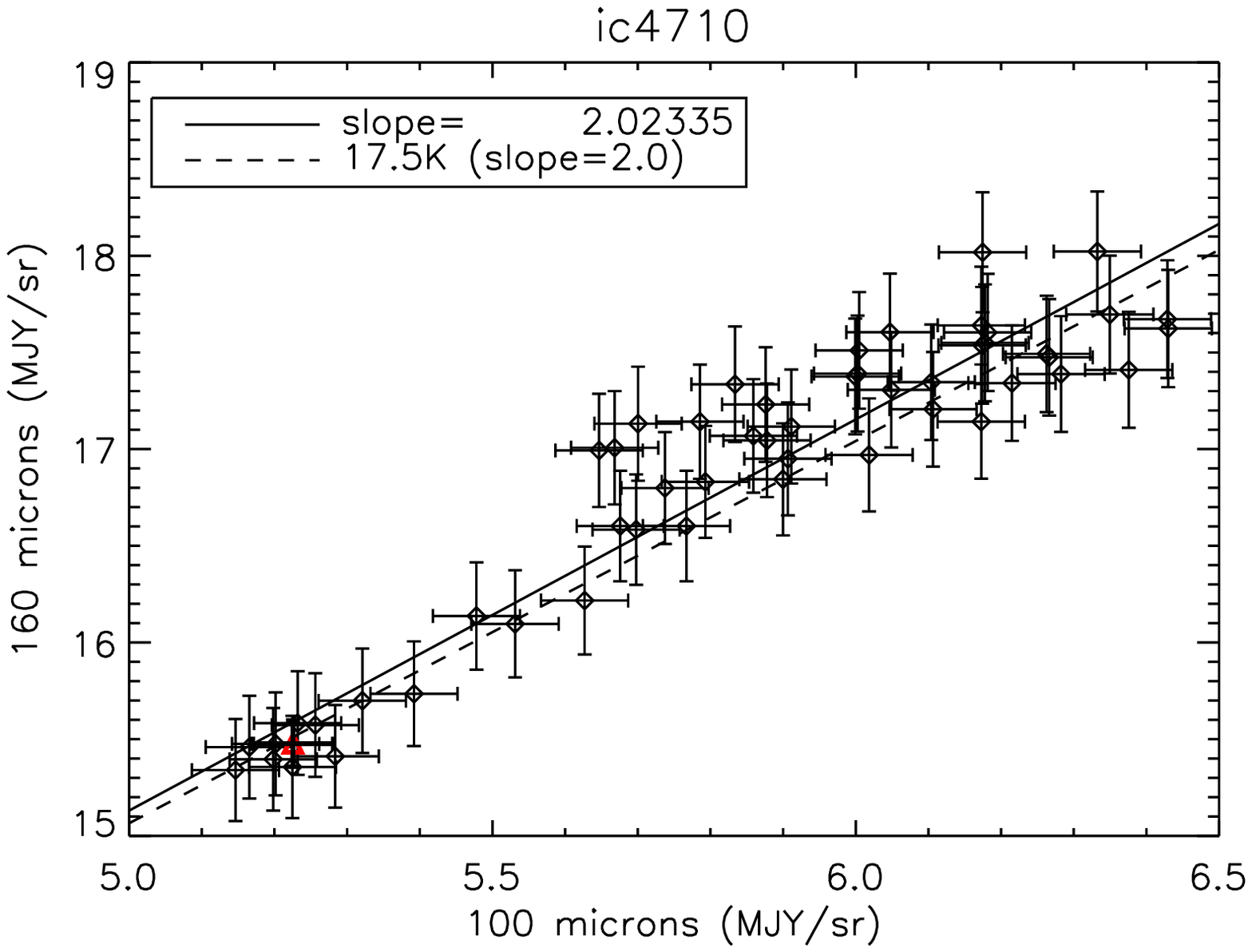}\\
\includegraphics[scale=.30]{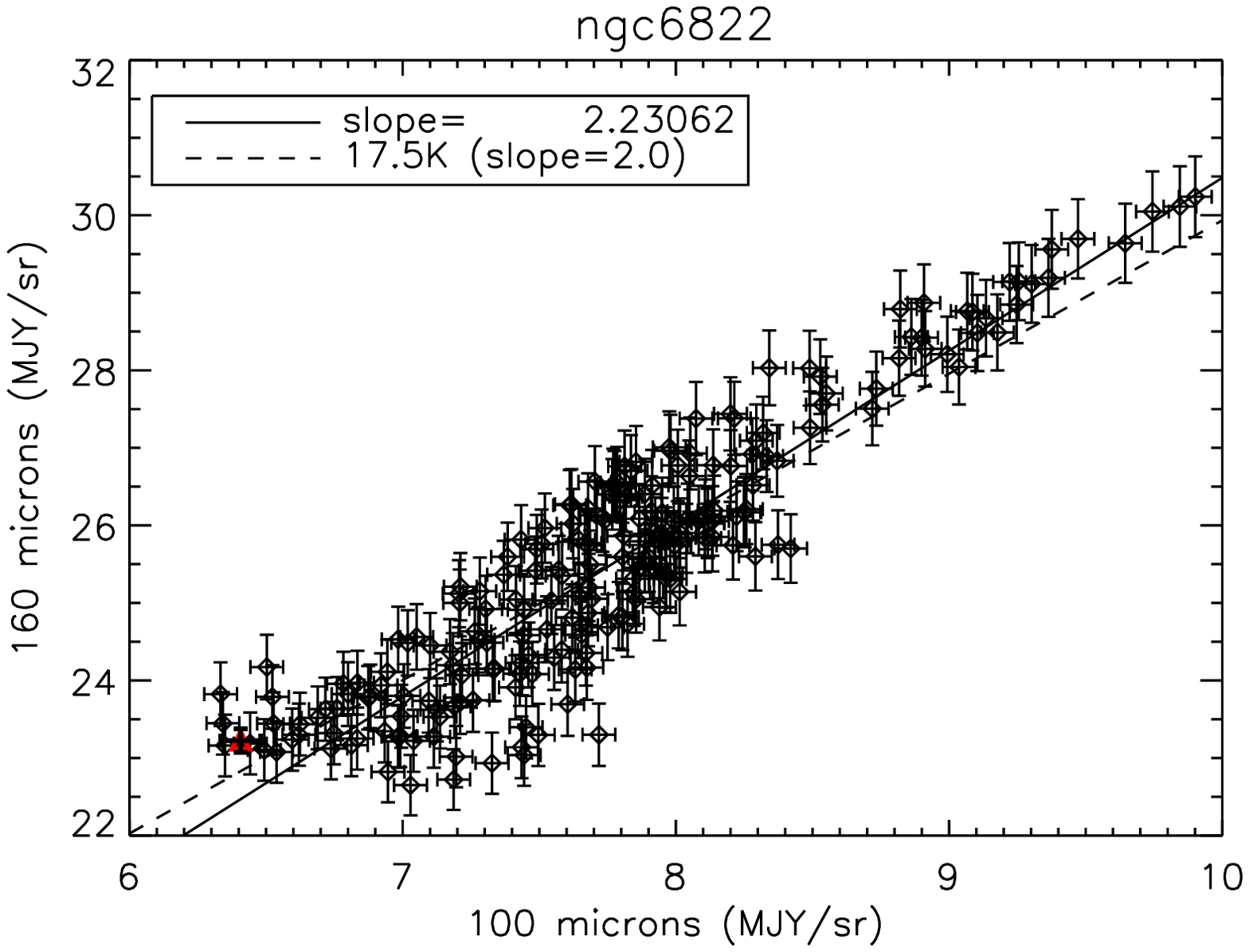}
\includegraphics[scale=.30]{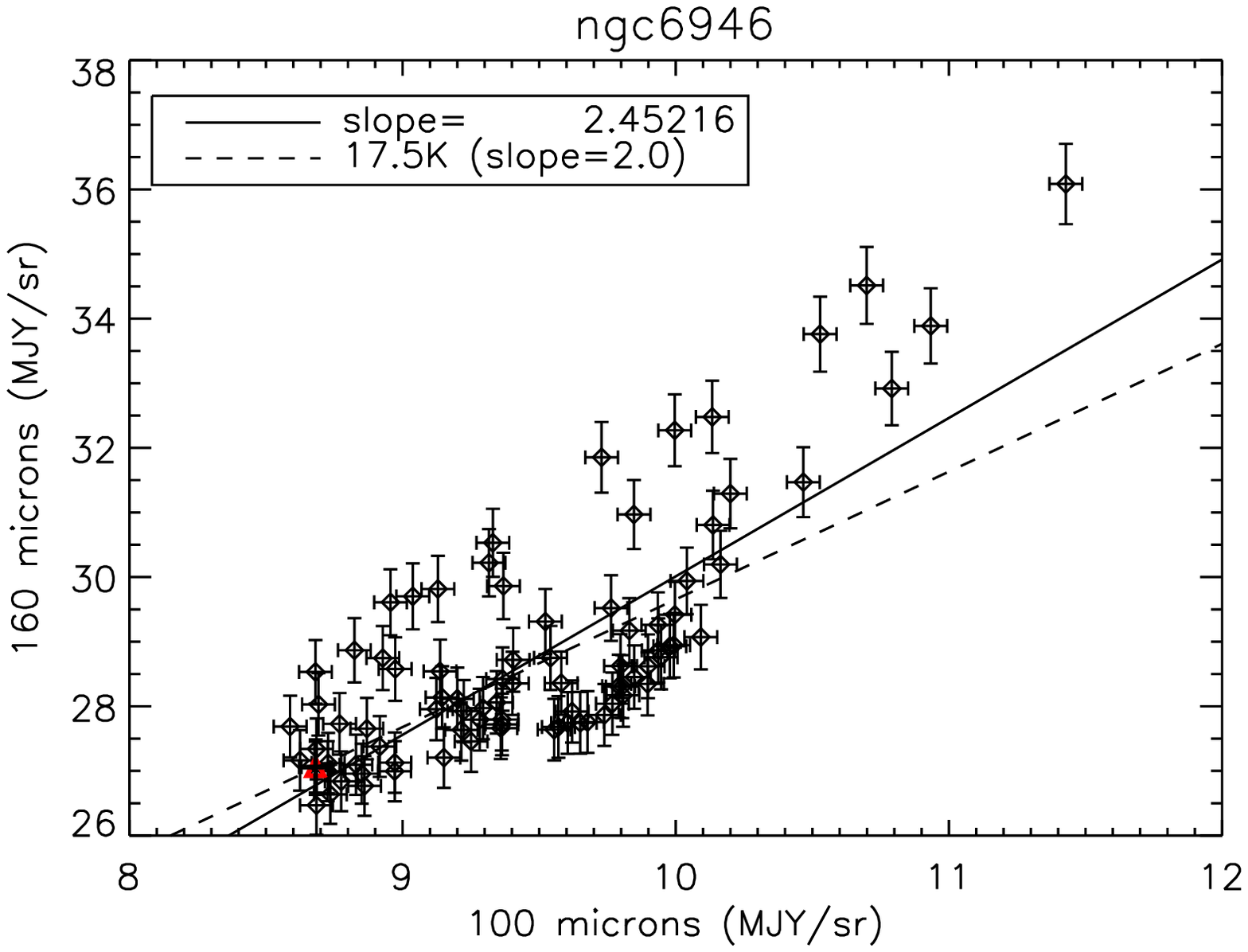}
\includegraphics[scale=.30]{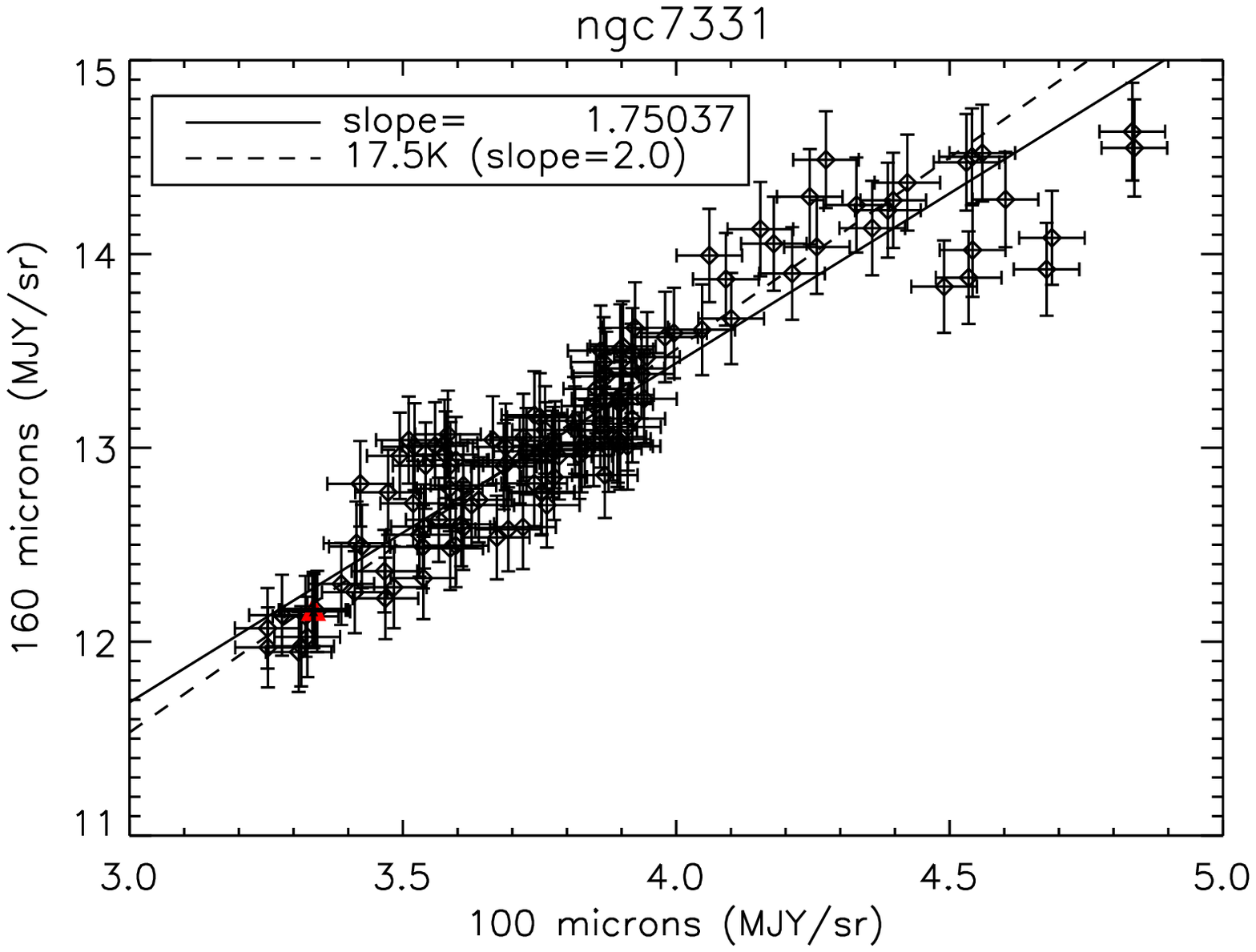}\\
\caption{ idem as Fig. \ref{fig2}\label{fig3}}
\end{figure*}

In each field of the selected sample, we plot the point to point correlation between the brightnesses observed at 100 and 160$\mu$m (represented in Fig \ref{fig2} and \ref{fig3}) and apply a linear fit taking into account the errors at both wavelengths. This enables us to obtain for each field a slope corresponding to the ratio $B_{160}/B_{100}$ unbiased by variations of the zero point level (residuals from the zodiacal light subtraction, absolute value of the CIB).  The correlation coefficient and the slope derived in each region are summarized in Tab. \ref{tab2}. 

Large scale observations of high galactic latitude emission of cirrus with COBE were well characterized by a modified black body with a dust temperature of 17.5K and an emissivity index proportionnal to $\nu^2$ \citep{BAB+96}. Using this law, we estimate the large scale 160/100 color for cirrus to be of $B_{160}/B_{100}=2.0$ (taking into account color corrections). This ratio is represented in the correlation plots (Fig. \ref{fig2} and \ref{fig3}) to guide the eye. While some correlations between $B_{100}$ and $B_{160}$ are in agreement with the $B_{160}/B_{100}=2.0$ obtained on large scales, clear deviations are also seen (4 fields out of 15 have a slope discrepant at a 5-6$\sigma$ level with respect to the value of 2.0. The most extreme case is the field of NGC2976, with a fitted slope on the $B_{160}$ versus $B_{100}$ correlation that is 10$\sigma$ away from the 2.0 standard value). 

\begin{figure}
\plotone{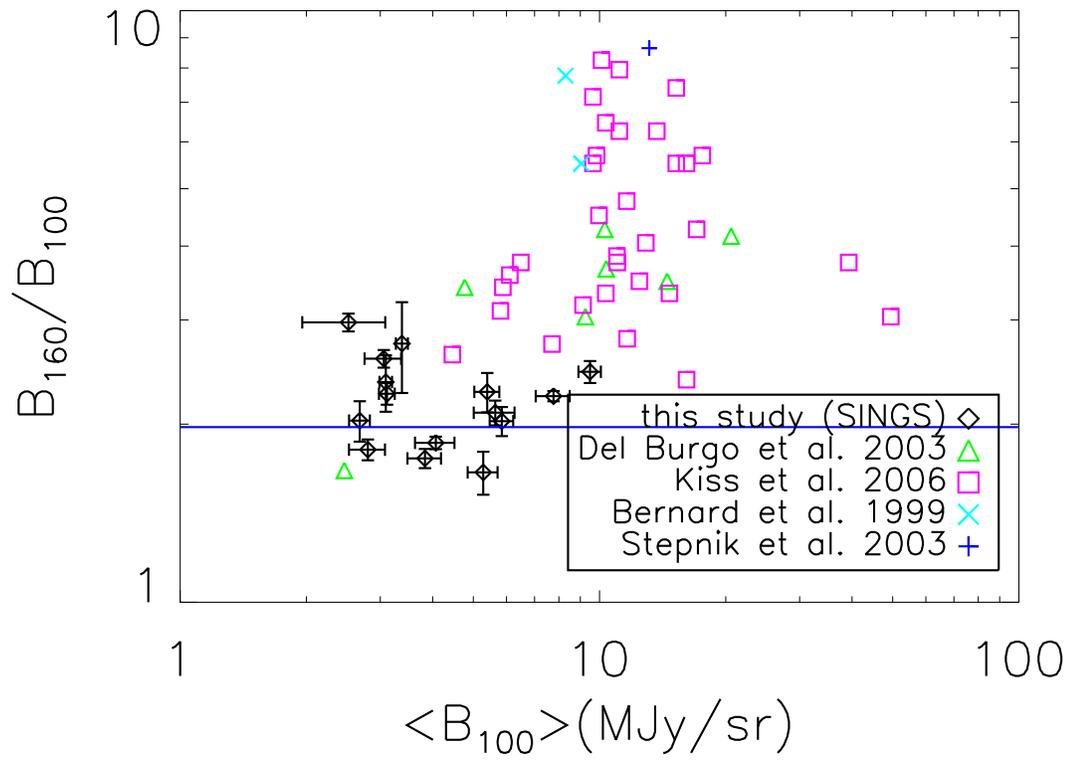}
\caption{Variations of the 160/100 surface brightness ratio with the mean 100$\mu$m surface brightness. The diamonds represent the SINGS observations. The blue line denotes a typical ratio of 2.0.\label{fig5}}
\end{figure}

In Fig. \ref{fig5}, we compare the obtained ratios $B_{160}/B_{100}$ to the mean surface brightness at 100$\mu$m in each field (black points). We observe a large dispersion in the 160/100 colors that can not be explained by the error on the data or the fitting process.  At 100 and 160$\mu$m, the interstellar emission is dominated by the emission from big dust grains at thermal equilibrium with the radiation field \citep{DBP90} and the $B_{160}/B_{100}$ ratio is therefore related to that characteristic dust temperature. Taking a standard emissivity of dust per hydrogen atom in H{\sc i} from \citet{BAB+96} and an emissivity index of 2, the 160/100 color variations we are probing can therefore be related for illustrative purposes to temperatures between 15.7 and 18.9K for column densities ranging from $N_H=3\times 10^{20}$ to $2\times 10^{21}$cm$^{-2}$. We note that these variations are consistent \emph{on average} with the large scale estimate (blue solid line), confirming that the fields used in this study are sampling the cirrus emission observed on large scale. 

For a given grain size and composition, this characteristic temperature depends on the local radiation field strength and spectrum which depends on the presence and distance of nearby heating sources and on the extinction.  In the framework of this model, the presence of large variations in the 160/100$\mu$m ratio observed in our sample would suggests the presence of large variations in the heating of grains at small scales (variations by a factor of 3 of the intensity of the incident radiation field). This can be surprising since at low FIR surface brightness and at high latitude the interstellar radiation field might be expected to be homogeneous.  We looked at the far-infrared color temperature maps derived from DIRBE observations by \citet{LABP98} and \citet{Schlegel:1998ys}. The regions we are studying appear to be reasonably representative of the high latitude cirrus given the small number statistics. For the sightlines covered by our sample, the FIR color variations seen in the DIRBE data are compatible with the variations that we observe. Our study is indeed more sensitive than previous observations and therefore able to probe color variations smaller than the uncertainties in the previous studies.

The shape of the optical spectrum heating the grains could also affect the far-infrared colors: the radiation field could become gradually harder with position off the galactic plane \citep{Mattila:1980zl}. Using the cirrus model from \citet{Efstathiou:2003xe} with different stellar populations heating the clouds, we checked that changes in the shape of the optical radiation field is unlikely to affect the 160/100 and 60/100 colors of cirrus by more than 20\%. 

The dust equilibrium temperature depends however also on the structure of the grains. Grain aggregates for example cool more efficiently. The temperature variations observed in the diffuse medium could therefore be either due to variations of the intensity of the interstellar radiation field or to changes in the grain structure.

We compared our findings with different studies of far infrared emission from the literature: the quiescent high galactic latitude clouds from \citet{del-Burgo:2003rt}, 
 the large sample from archival ISOPHOT data by \citet{Kiss:2006uq}, the two regions in a high latitude cirrus MCLD 123.5+24.9 observed by \citet{Bernard:1999qy},
and the quiescent filament in the Taurus molecular complex from \citet{SAB+03}.
 Because other observations were obtained with different instruments, we have to interpolate the brightnesses at various far infrared wavelengths to estimates at 100 and 160$\mu$m. To do so, we took the dust temperatures determined in each study with a brightness at 100 or 200$\mu$m and used a modified black body law with a spectral index. The power index is either taken from the study itself (if it was computed) or is fixed to a standard value of 2. For each region, we also compute the mean 100$\mu$m brightness as observed by IRIS and subtract a mean CIB contribution as for our observations. Despite  large scatter, we observe a trend between $<B_{100}>$ and the 160/100 color that is consistent with the idea that denser regions are colder. However, the effect of selection biases of these studies remains unclear. The comparison of our results with that from the literature (Fig. \ref{fig5}) shows that previous studies in the far-infrared have been targeting higher $B_{160}/B_{100}$ and $<B_{100}>$, i.e. denser and colder clouds. Due to our better sensitivity, our observations fill the gap at low $B_{100}$ and low $B_{160}/B_{100}$. 
  
For the first time, we observe interstellar dust emission at low surface brightness in an unbiased way (in the observing strategy) with a high sensitivity. These observations show that former studies on dust properties at FIR wavelength at small scale, have been biased toward colder and denser clouds. Our study shows that the 160/100 brightness ratios of high galactic cirrus clouds at small scales are consistent on average with the observations on large scales. However, these 160/100 colors show a wide dispersion that could arise from variations in the heating of the clouds or from change of the dust grain structure. In order to investigate further the origin of the 160/100 variations, we extend the comparison to the 60$\mu$m data.

\subsection{Comparison to the 60$\mu$m data}

\begin{figure*}
\includegraphics[scale=.30]{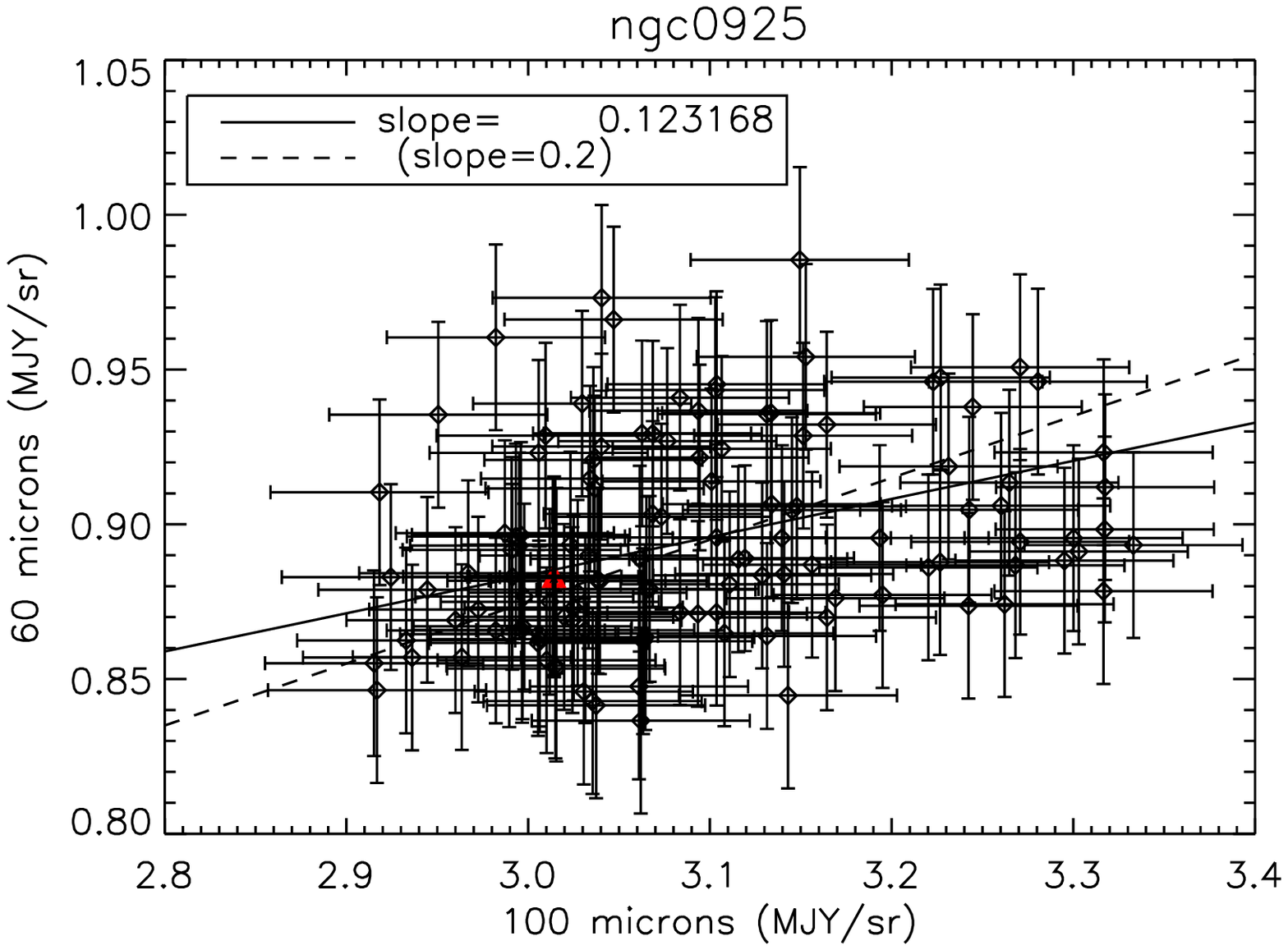}
\includegraphics[scale=.30]{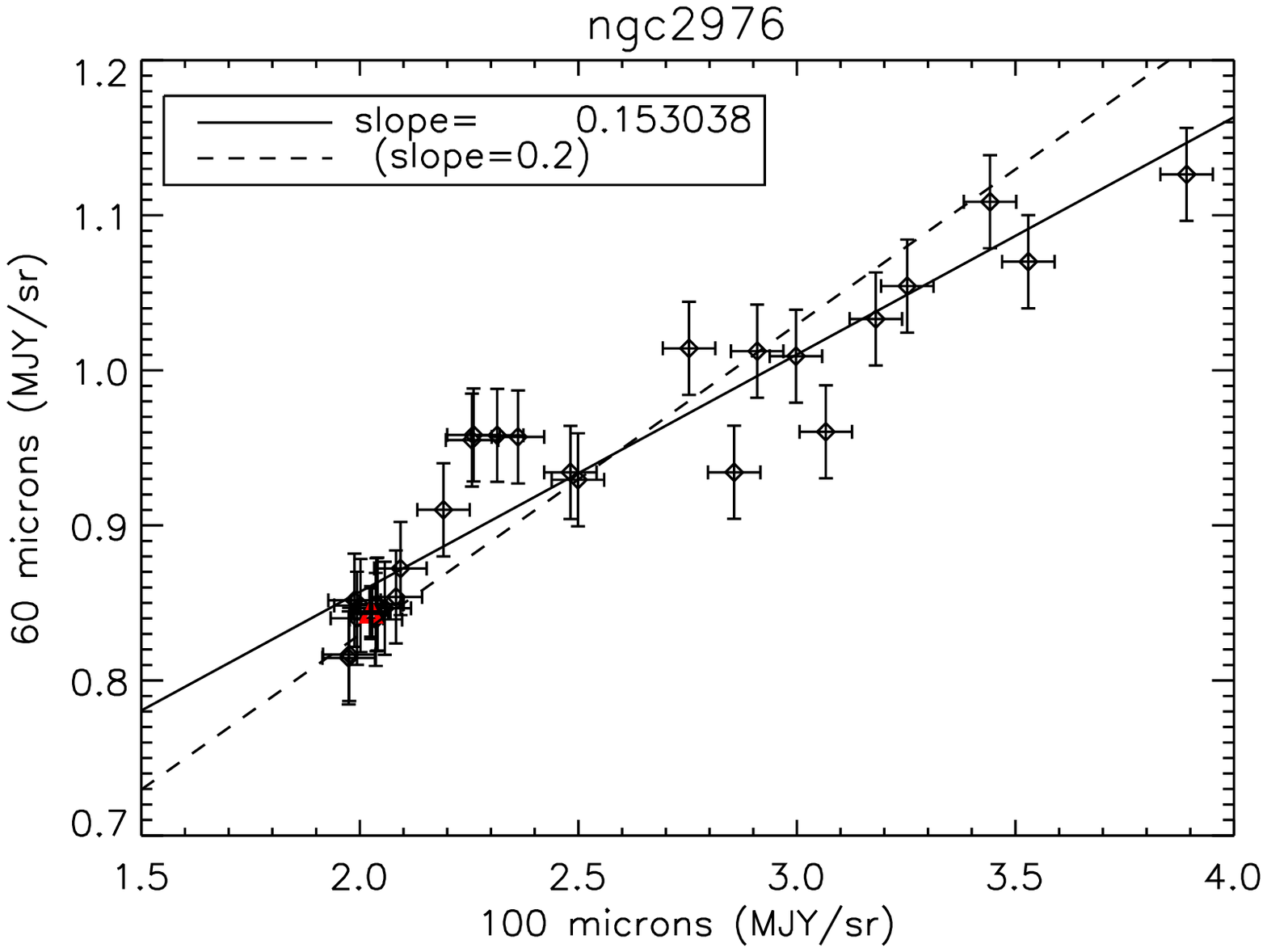}
\includegraphics[scale=.30]{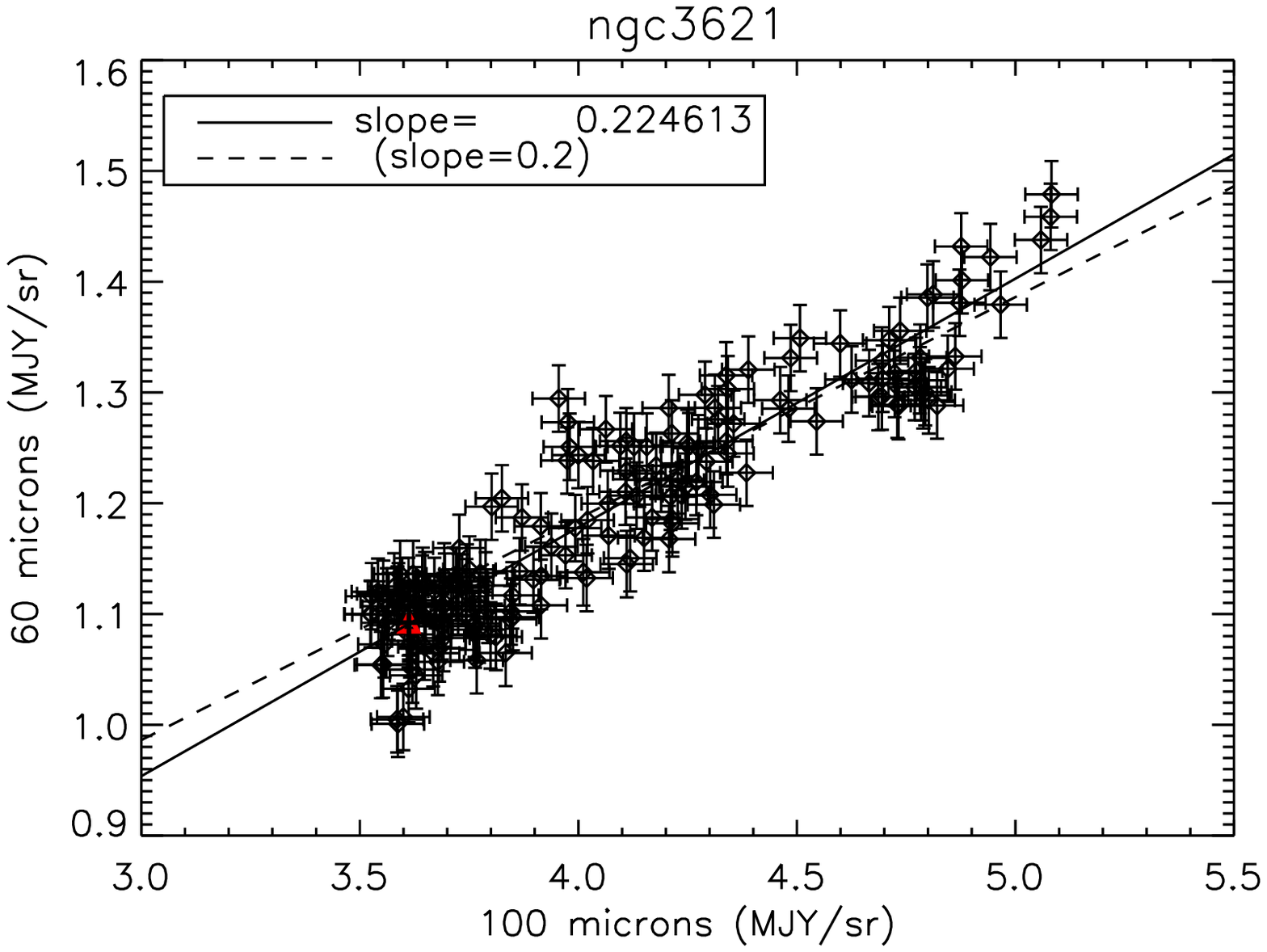}\\
\includegraphics[scale=.30]{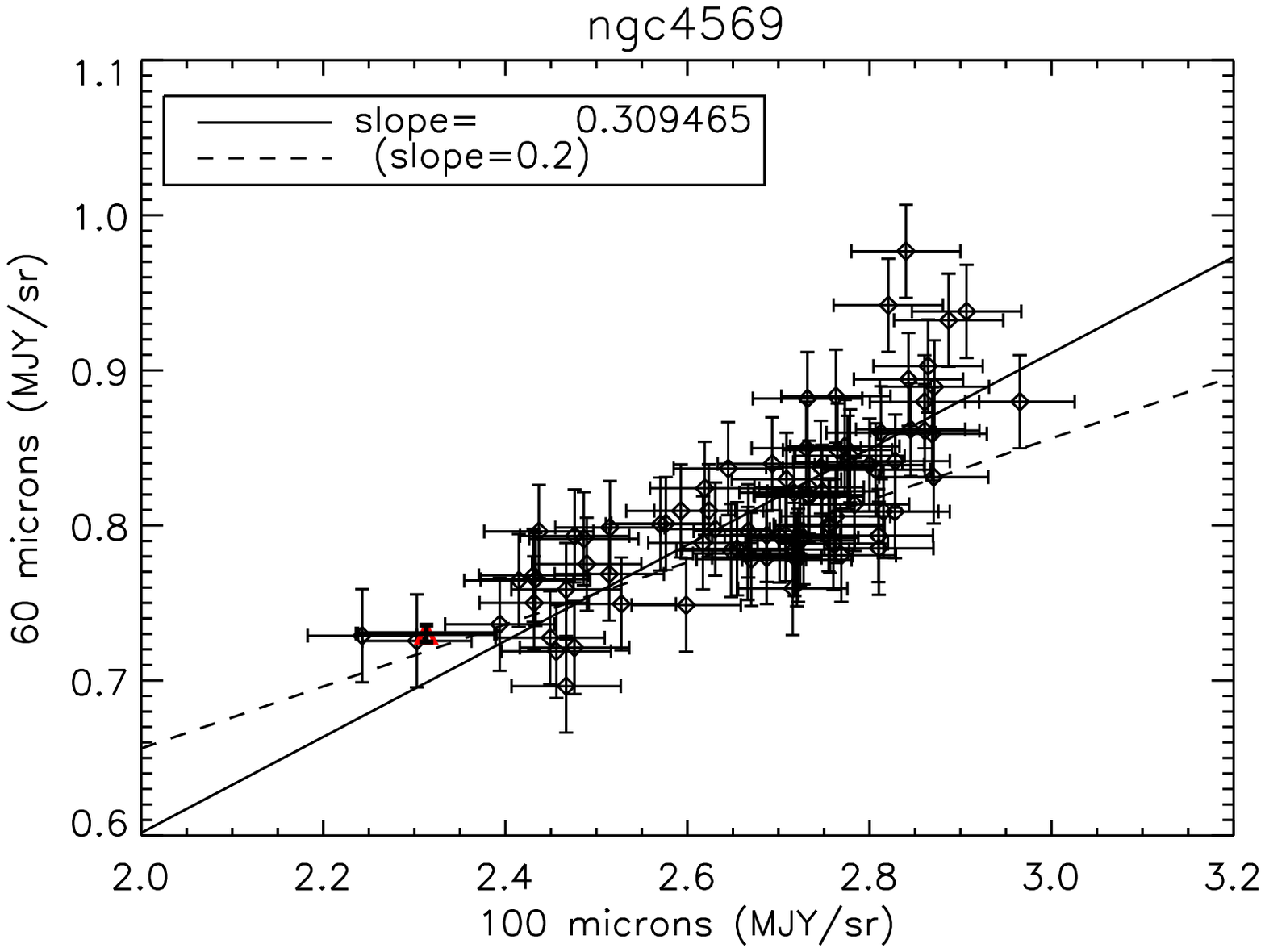}
\includegraphics[scale=.30]{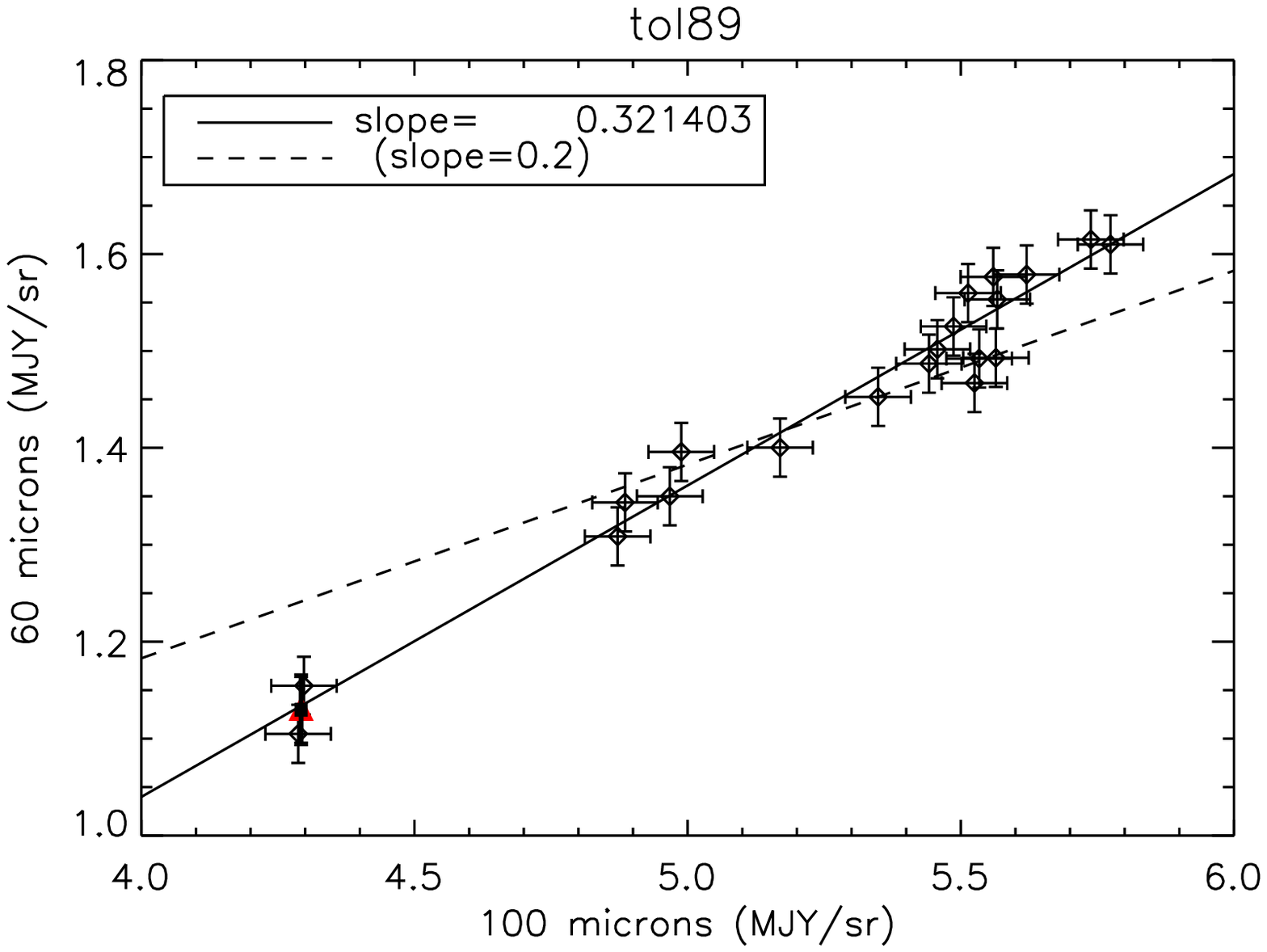}
\includegraphics[scale=.30]{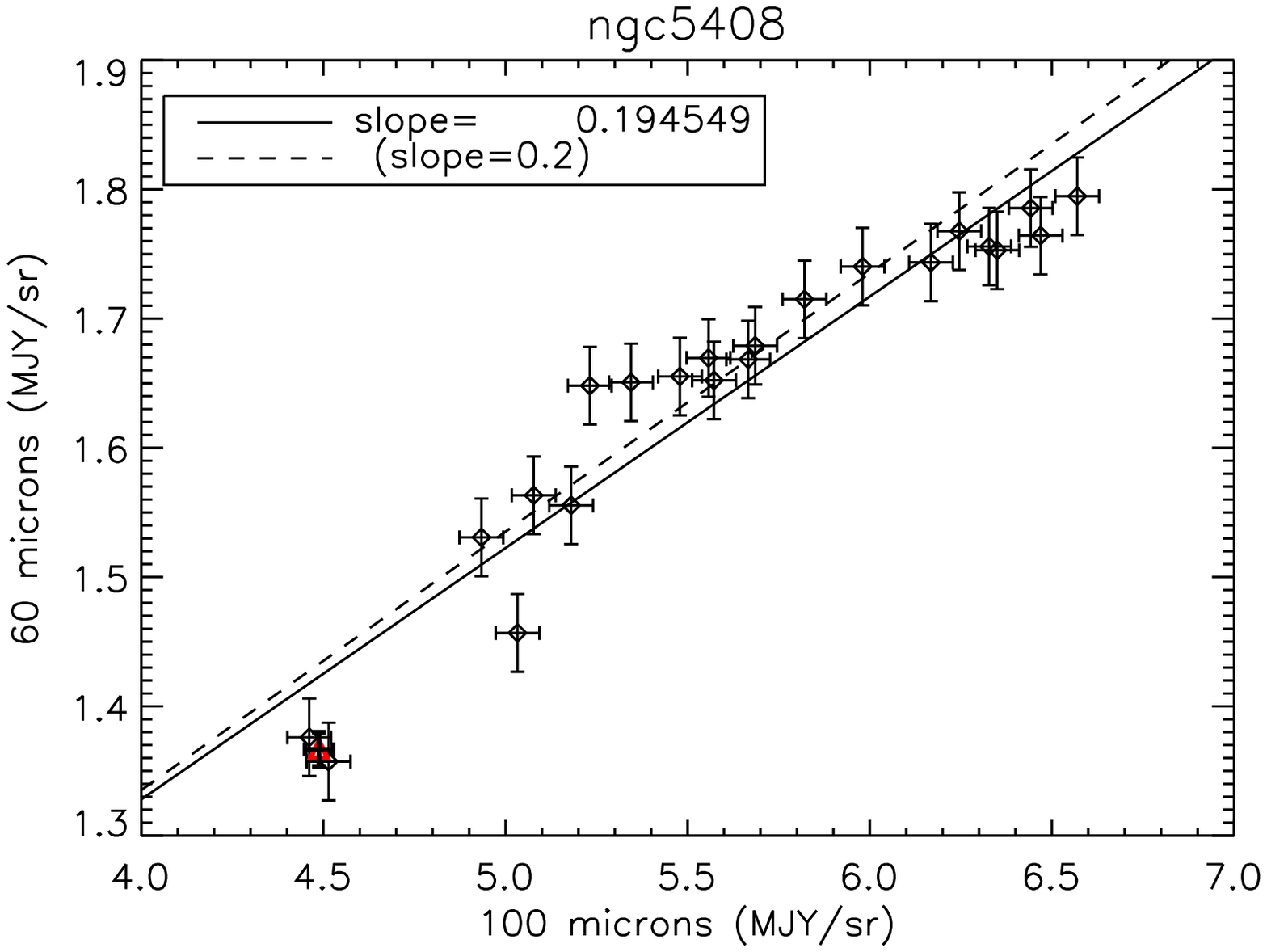}\\
\includegraphics[scale=.30]{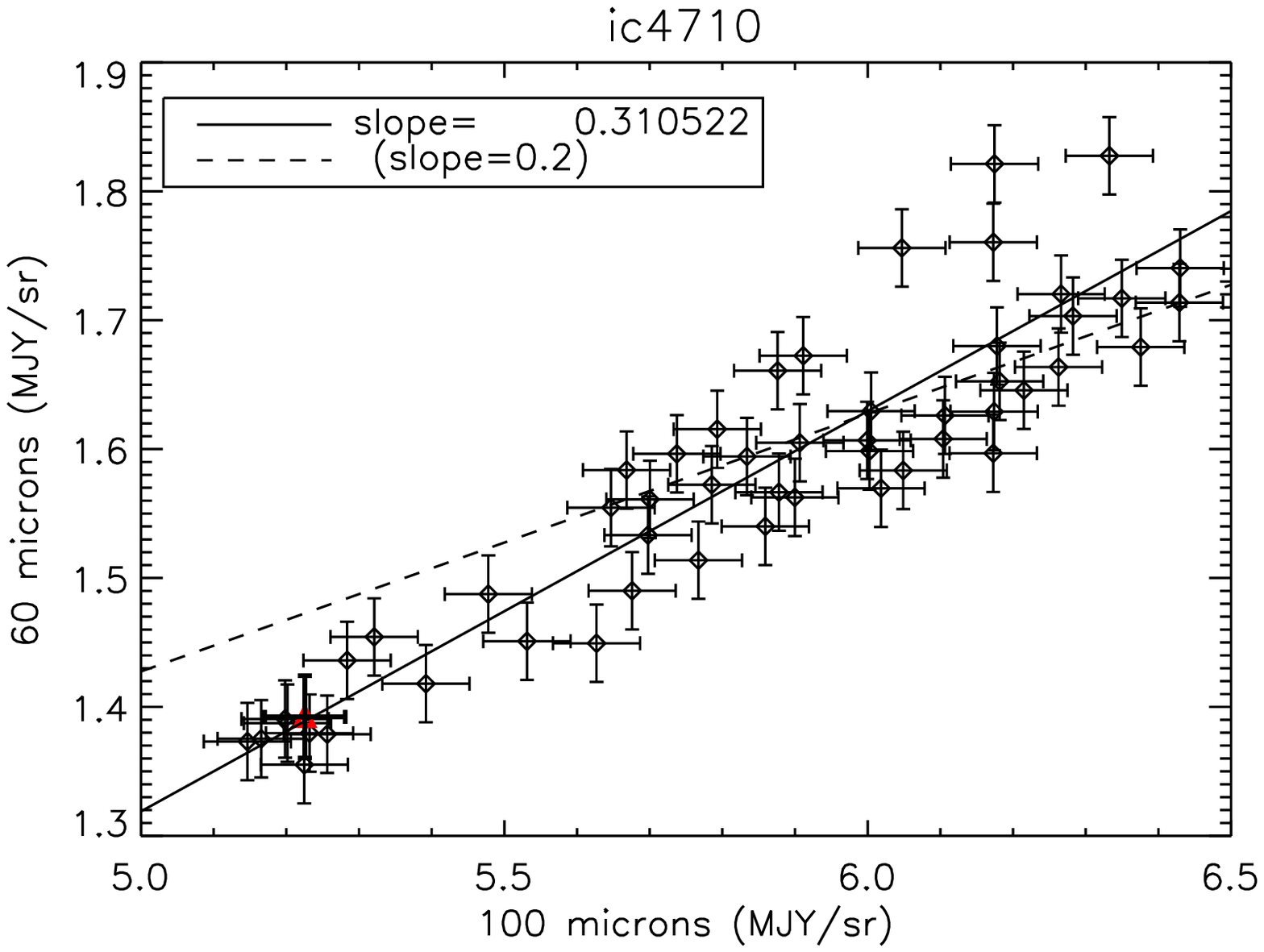}
\includegraphics[scale=.30]{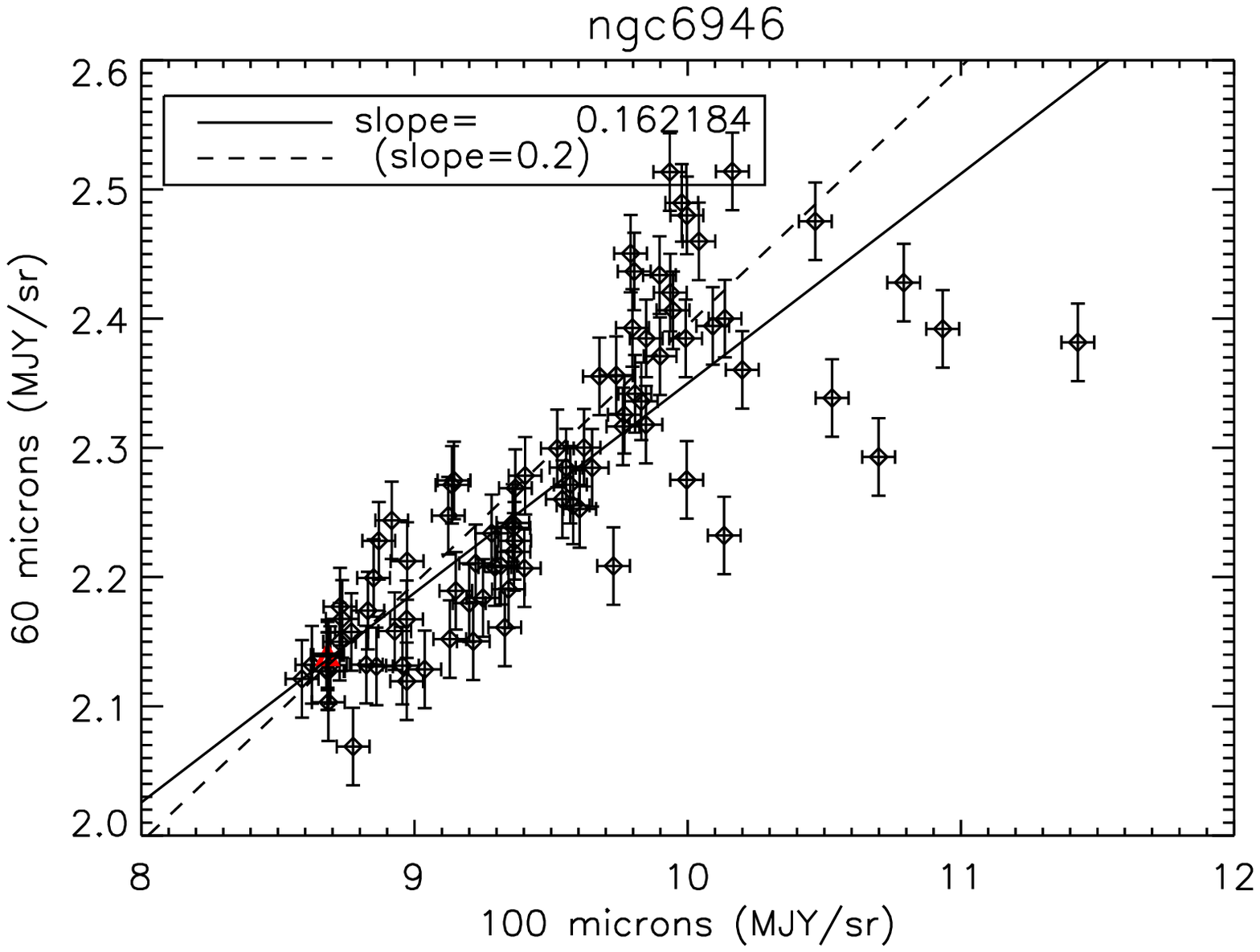}
\includegraphics[scale=.30]{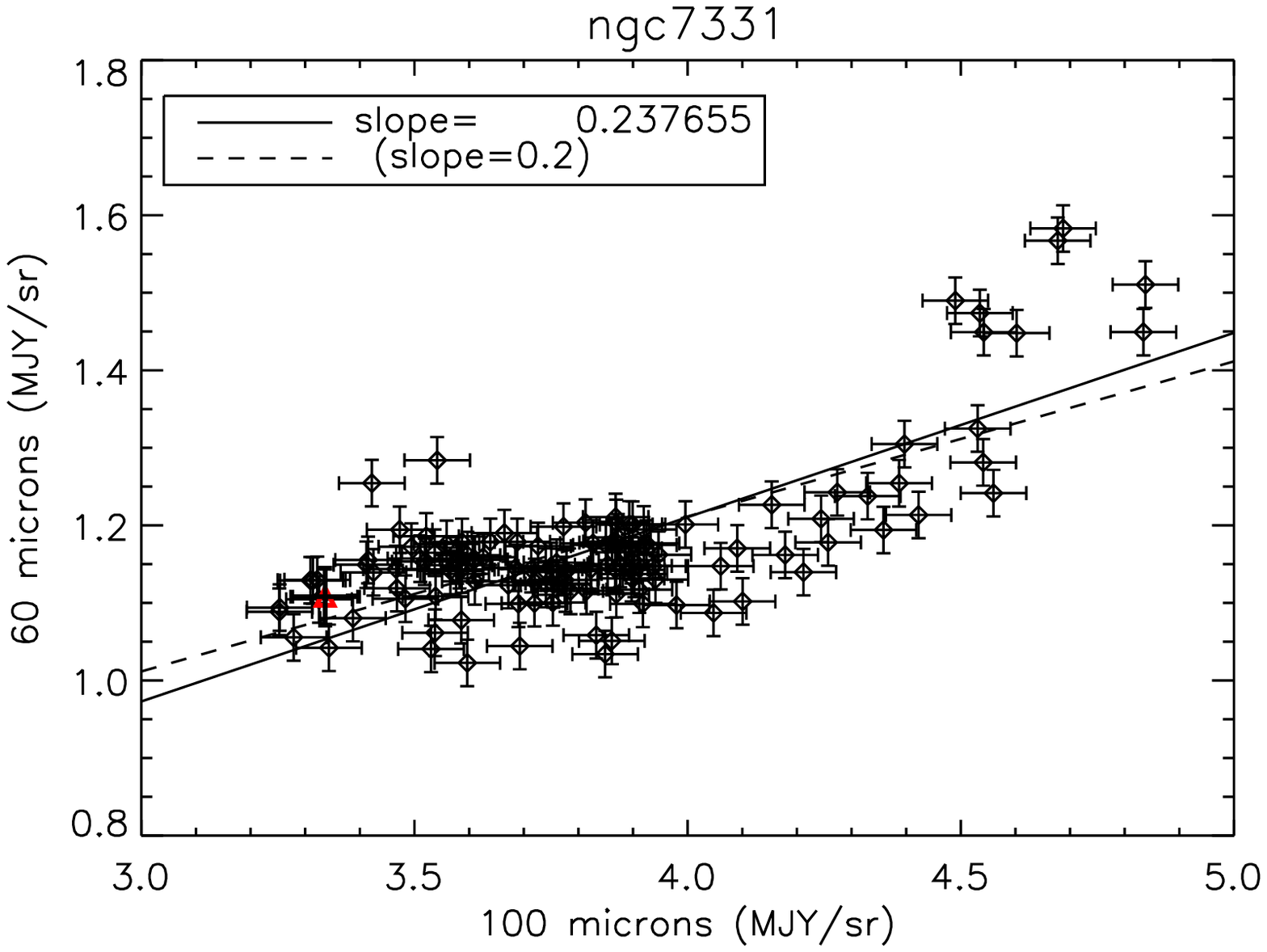}
\caption{ 60--100 scatterplots for all SINGS observations with $<B_{100}> \geq 2.5$ MJy/sr. In each plot, a canonical slope of 0.2 is represented by a dashed line. A linear fit is applied and the best fit is represented with a solid line. The value of the slope obtained is written in the legend. The green point and its bold uncertainties represent the value chosen to represent the local background in the observation.\label{fig4}}
\end{figure*}

To investigate the origin of the 160/100 color variations observed in the diffuse cirrus, we compare the 160 and 100$\mu$m data to the 60$\mu$m emission. The sample for this part of the study is however reduced to fields with an ecliptic latitude above 15$^o$ in order to avoid artifacts due to the uncertainties in the zodiacal light subtraction in the IRIS data. For each field, we determine a 60/100 color by using the same correlation technique as above. The correlations in each region are shown in Fig. \ref{fig4} and the obtained $B_{60}/B_{100}$ are summarized in Tab. \ref{tab2}.

\begin{figure}
\plotone{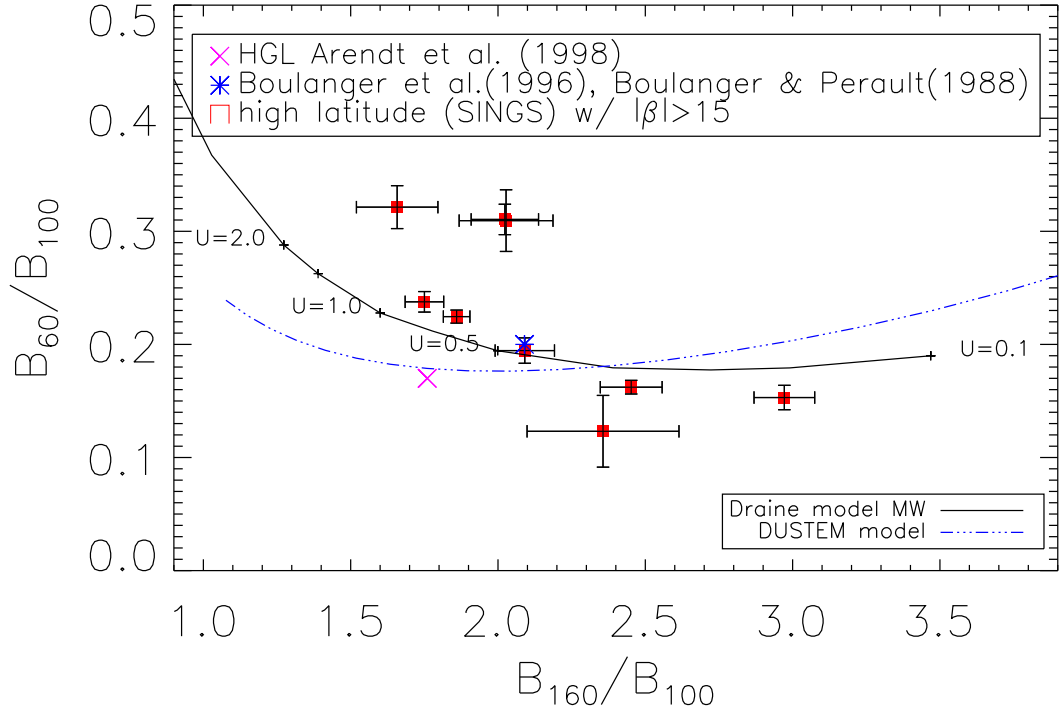}
\caption{Color-color plot showing the evolution of the 160/100 and 60/100 brightness ratios with each other for all the the regions where $<B_{100}> \geq 2.5$ MJy/sr. Two reference values from the literature representing the colors measured on large scales are overplotted. The observed colors are compared to the trend obtained with two dust models: the Draine et al. (2007) model (dark solid line) and the DUSTEM model (blue dotted line). In both case, the incident interstellar radiation field intensity is the only parameter varying (the proportion of grain types are kept constant). \label{fig6}}
\end{figure}

Fig. \ref{fig6} presents the $B_{60}/B_{100}$ ratio obtained in each field with respect to the $B_{160}/B_{100}$ ratio. Here again,  large variations are observed in the 60/100 colors  that can not be explained by the uncertainties in the data or in the analysis. As for the 160/100 colors, the 60/100 brightness ratios are consistent on average  with the "reference values" (the pink cross and the blue star in the figure) obtained for high latitude emission on large scales \citep{BP88,BAB+96,AOW+98}. Furthermore, there is a trend of decreasing $B_{60}/B_{100}$ with increasing $B_{160}/B_{100}$. 

To try to interpret this trend, we used two models of the dust grain emission at different interstellar radiation fields: the \citet{Draine:2007lr} model for the Milky Way \footnote{We took the model with a PAH fraction q$_{PAH}=4.58\%$ but we checked that this parameter does not influence significantly the results of this study} and the "DUSTEM" model (updated model based on \citet{DBP90}). The models take into account the shape of the IRAS and MIPS/Spitzer filters, the color corrections. For both models, the abundances of different grain types are kept constant. The tracks obtained are compared to the data in Fig. \ref{fig6}. The comparison shows that, if the variations of colors are due to variations in heating of the grains, this would imply large changes in the interstellar radiation field at high galactic latitude (from $U\approx 0.3$ to 1). Furthermore, the trend observed between the 60/100 and 160/100 colors is not well reproduced with the current models by changing the interstellar radiation field alone. In that respect, the  \citet{Draine:2007lr} model is however closer to the observed trend than the DUSTEM model (only 3 fields are more than 3$\sigma$ deviant from the expected curve), but for $B_{160}/B_{100}<2.1$ all observations show systematically higher $B_{60}/B_{100}$ than predicted by both models, while for $B_{160}/B_{100}>2.1$ all data points have systematically lower $B_{60}/B_{100}$ values than expected from both models. 

Current dust models might be missing an additional dust grain type. Such an addition might reproduce all color variations while changing the interstellar radiation field alone.
Another way to interpret the observed trend is that the variations in the dust emission spectrum reflect spatial changes in the grain properties -- composition, structure or size distribution. 

The equilibrium temperatures of dust grains is expected to decrease for increasing grain sizes and small grains ($\leq 0.01\mu$m) undergo temperature excursions following single-photon heating that enhances the 60$\mu$m emission. Thus regions with fewer small grains may have lower 60/100 ratios. The observed trend between the 60/100 and 160/100 infrared colors could be reproduced by changing the dust grains size distribution or composition. For example, enhancing the amount of small grains  in regions with higher interstellar radiation field (i.e. higher temperatures) and reducing it at low equilibrium temperatures could reproduce the observed variations.  

In the same way, regions with enhanced populations of large grains may have increased 160/100 ratios. In that case, reproducing the observations could be obtained with only modest variations in the starlight heating rate and shifts in the grain size distribution (fewer small grains and increased sizes for the larger grains at low temperatures, more small grains and smaller sizes for the big grains at higher dust temperatures). 

The 60/100 colors we observe therefore suggest changes of the dust properties (dust size distribution and/or composition) from one region to the other. These changes are related to variations in the 160/100 brightness ratio. Whether the 160/100 color variations require a change of the starlight intensity  heating the clouds or result from the change of dust properties alone is unclear. 
The interpretation of the color variations and of the trend between the 160/100 and 60/100 colors is discussed further in the next section.

\section{Discussion\label{sec:discuss}}

Despite its rather constant color distribution on large scale, the far infrared emission from diffuse cirrus at high galactic latitude is observed to host large color variations on small scales. These variations seem related to each other (the 60/100 color decreases as the 160/100 color increases). In this section, we will first check that these variations come indeed from cirrus emission and are not related to the galaxies targeted with the observations. Second, we will discuss possible interpretations for these large color variations and the trend between them.

\subsection{Extended disks in galaxies}

Because the MIPS observations were taken to observe nearby galaxies, it is legitimate to ask whether the infrared emission that we observe could be associated with these targets. In particular, H{\sc i} observations have shown that gaseous disks can extend much farther than the optical diameters. Dust grains could be present in these outer parts of the galaxies and bias our measurements. 

To avoid this extended emission from the galaxies, we were careful in masking regions larger than the detected emission (c.f. Sec. \ref{sec:dattreat}). Some of the galaxies in the observations used in this study have been observed in H{\sc i} observations through The HI Nearby Galaxies Survey  \citep[THINGS,][]{Walter:2005fj}. We checked that the H{\sc i} diameters reported for these galaxies are smaller than the region masked for our study. We are therefore confident that the variations observed in the infrared emission between fields do not come from the targeted galaxies, but rather from diffuse cirrus emission.

\subsection{Physical conditions in cirrus clouds}

A possibility to interpret the variations of far-infrared colors at high galactic latitude is that we are sampling clouds in different physical conditions and/or composition (different heating of the clouds, different dust size distribution, \ldots). The color changes would then be due to mixing along the sightline of these different components. 

An unbiased survey of H$_2$ absorption in high galactic latitude clouds by FUSE  \citep{Gillmon:2006rm} has been interpreted as showing that some clouds have been compressed. The dynamical history leading to this compression may involve shock waves or strong turbulence, which could also lead to changes in the grain size distribution by shattering in grain-grain collisions, possibly explaining the regions of higher than average 60/100 and lower than average 160/100 colors.  

One tantalizing possibility, in terms of mixing, is the presence of Intermediate or high velocity clouds (IVCs and HVCs) along the sightline. These cloud falling onto our galaxy could have very different dust properties \citep[e.g.][]{Miville-Deschenes:2005cf} and would bias the measured infrared emission from more local cirrus clouds.
  We checked for the presence of intermediate velocity clouds in the LAB H{\sc i} survey spectra \citep{Kalberla:2005ds} in the direction of the fields in our sample. For about half of the sample, there is a intermediate velocity component seen in H{\sc i} in the sightline. Two sightlines also have a high velocity component. However, no conclusive trend between the fraction of the H{\sc i} in the IVCs and/or HVCs and the infrared colors could be seen. This may be due to the lack of resolution of the H{\sc i} observations ($\sim 0.6^o$), to the difficulty to disentangle IVCs and the Milky Way in some regions or to the small size of our sample.

\subsection{Variations in grain properties}

 The grain size distribution is the result of processes such as sputtering, shattering and coagulation, and sightline-to-sightline variations in the wavelength-dependence of optical and ultraviolet extinction toward stars have already demonstrated regional variations in the grain properties. Whether the observed variations in emission can be fully explained by variations in the size-distribution alone, or whether other properties (e.g., composition or porosity) are also involved is uncertain.

In denser clouds, variations of infrared colors \citep{SAB+03,Kiss:2006uq} have been interpreted with a grain coagulation scenario, combining changes of the size distribution with changes of the dust emissivity properties. The trend observed between the 60/100 and 160/100 colors would be consistent with this idea. In this scenario, most of the 160/100 color variations would then be due to the change of emissivity of dust grains (due to changes of their structure), while the 60/100 colors would change with the incorporation or release of small grains in large dust aggregates. We could be witnessing variations of dust properties due to variations in turbulent motions in the diffuse interstellar medium. Alternatively, dust grains in the diffuse medium could retain for some time the aggregate structure they had previously acquired in denser regions. So we could be seeing a sequence of regions corresponding to increasing time since their release from high density environments.

Such changes in the dust size distribution and structure of grains would imply related variations of the UV-optical extinction curves at high galactic latitude. An extinction and reddening study of stars at high galactic latitude behind translucent clouds \citep{Larson:2005ao} shows variations in the extinction curves obtained with respect to the average curve for the diffuse interstellar medium. In particular, 48\% of their sightlines have $R_V<2.8$, much lower that the diffuse ISM average of $3.05$. Such values are indicative of enhanced abundances of small grains, and these regions could have a higher that average 60/100 color. To test if the high 60/100 colors and low $R_V$ values are connected, we computed the 60/100 colors from IRIS data in a similar fashion to this study for a $\sim 1\times1^o$ region around each sightline of the \citet{Larson:2005ao} sample. We do not observe any correlation however between the $R_V$ and 60/100 color, nor between $A_V$ and the 60/100 color. Unfortunately, no longer wavelength observations exist for these regions and it would be important to determine the 160/100 colors in these regions as well and study their dependancy with extinction properties. This result is however a concern for the coagulation scenario as an interpretation of the 60/100 color variations we observe.

Interpreting the trend observed between the 60/100 and 160/100 colors with a change of dust optical properties and dust size distribution remains hypothetical without further observations. In particular, H{\sc i} observations of these diffuse sightlines, with a high resolution (at least similar to the IRIS one), will be needed to determine the emissivity of dust per hydrogen atom and test if variations are observed and correlated with the infrared colors. 

It is presently not possible to interpret further the far-infrared color variations in terms of physical condition changes, grain size distribution, grain properties, etc. A larger number of observations of high latitude cirrus could help to probe the spatial variations. 
H{\sc i} studies of these regions at high resolution would also be crucial to probe possible variations of dust grain emissivities, or to check whether the velocity structure of the cirrus correlates with the 60/100/160 colors. Finally, extinction curves on sightlines where cirrus emission properties have been determined would be most useful to see whether extinction properties would correlate with the FIR colors. 

\section{Conclusion}

We performed an unbiased study of dust emission from high galactic latitude cirrus using serendipitous Spitzer MIPS observations at 160$\mu$m from the SINGS survey, complemented by IRIS data at 60 and 100$\mu$m. After an appropriate post-reduction of the data and a removal of the targeted galaxy, a sub-sample is selected so that the variations of the cirrus emission dominate over the CIB fluctuations in each field.

We observe 160/100 colors in our fields that are consistent on average with large scale studies. However, strong variations are also observed from field to field. This paper extends former studies on dust properties at high galactic latitude to more diffuse, fainter and warmer clouds. The 60/100 color is also observed to vary significantly in the sample and there is a trend of decreasing 60/100 with increasing 160/100 ratios. This trend is not completely reproduced by current models taking only into account variations of the radiation field strength and requires changes in the dust properties, composition or size distribution.

The exact origin of these variations remains unknown, but the variations of the 60/100/160 colors may reflect variations of the grains size distribution, of grain properties in addition to heating variations. However, we can not completely rule out the possibility that our fields contain emission from matter at different heights above the Galactic plane, the juxtaposition of multiple components in the fields could be affecting the infrared color estimates.

All in all, we observe unexpected variations of far-infrared colors in the supposed "homogeneous" cirrus at high galactic latitudes. These variations are not yet understood and further studies will be needed to test their origin. In particular, studies on a larger area of sky is needed to confirm the significance of these variations and their spatial distribution on the sky could give new clues on their origin.  

These infrared color variations will most probably be linked with variations of the infrared colors at longer wavelengths. They therefore represent an important point to study for the Planck  and Herschel missions. Such longer wavelength observations will enable us to determine precisely the temperature and spectral index of the dust, and their variations, in high latitude regions.

\acknowledgments

We would like to thank our referee M. Rowan-Robinson for his useful and interesting comments and A. Efstathiou for his help with their cirrus model. We are also grateful to L. Cambr\'esy for useful discussions and work on upper limit for extinction in our fields.

{\it Facilities:} \facility{Spitzer}.

\bibliographystyle{aa}
\bibliography{../../biblio}

\clearpage

\clearpage

\clearpage

\clearpage

\begin{deluxetable}{c c c c c c c c}
\tabletypesize{\tiny}
\tablecaption{Characteristics of the observations: galactic and ecliptic coordinates, average 100$\mu$m cirrus brightnesses in each field as well as standard deviations at 60, 100 and 160$\mu$m\label{tab1}}
\tablewidth{0pt}
\tablehead{
\colhead{field} & \colhead{(l,b)} & \colhead{($\lambda$,$\beta$)} & \colhead{Area (deg$^2$)}&  \colhead{$\sigma_{60}$} & \colhead{$<B_{100}>$ (MJy/sr)} & \colhead{$\sigma_{100}$} & \colhead{$\sigma_{160}$} }
\startdata
NGC0337 & (      126.983 ,      -70.4576) & (      10.0714 ,      -12.6317) &     0.048 &            ---- &       5.75 &      0.713 &       1.951 \\
NGC0584 & (      148.863 ,      -68.1957) & (      17.8130 ,      -15.3840) &      0.100 &           0.057 &  2.38 &      0.177 &      0.387 \\
NGC0628 & (      138.043 ,      -46.3226) & (      27.4637 ,       5.05541) &      0.115 &             ---- &    3.09 &      0.287 &      0.809 \\
NGC0855 & (      143.962 ,      -32.1481) & (      39.9931 ,       13.3583) &     0.056 &            ---- &       3.33 &      0.111 &      0.399 \\
NGC0925 & (      144.527 ,      -25.8230) & (      44.7625 ,       17.7346) &      0.124 &           0.040 &  3.09 &      0.101 &      0.316 \\
NGC1097 & (      227.959 ,      -65.0380) & (      26.0183 ,      -43.9238) &      0.113 &           0.037 &  1.08 &      0.112 &      0.250 \\
NGC1291 & (      247.760 ,      -57.5207) & (      27.3597 ,      -55.9323) &      0.154 &           0.018 & 0.68 &      0.104 &      0.212 \\
NGC1316 & (      240.631 ,      -56.8186) & (      32.3476 ,      -53.2688) &      0.197 &           0.042 &   1.00 &     0.093 &      0.189 \\
NGC1377 & (      212.367 ,      -52.3728) & (      44.7413 ,      -38.7417) &     0.042 &          0.029 &   1.65 &     0.071 &      0.246 \\
NGC0024 & (      40.4333 ,      -80.1196) & (      351.074 ,      -23.9783) &     0.099 &          0.043 &   1.14 &     0.073 &      0.130 \\
NGC1404 & (      237.002 ,      -53.9016) & (      38.2431 ,      -52.8271) &     0.069 &          0.032 &  0.59 &     0.059 &      0.101 \\
NGC1482 & (      213.765 ,      -48.2291) & (      50.1099 ,      -39.5384) &     0.039 &          0.040 &   2.44 &     0.087 &      0.212 \\
NGC1512 & (      248.603 ,      -48.4042) & (      40.9466 ,      -61.8125) &      0.176 &           0.042 & 0.49 &     0.075 &      0.132 \\
NGC1566 & (      264.199 ,      -43.5133) & (      32.0129 ,      -73.2076) &      0.116 &           0.025 & 0.39 &     0.081 &      0.124 \\
NGC1705 & (      260.913 ,      -38.8932) & (      50.3989 ,      -74.3634) &     0.043 &          0.033 &  0.42 &     0.094 &      0.121 \\
NGC2403 & (      150.172 ,       28.7424) & (      102.746 ,       43.5117) &      0.261 &           0.049 &  1.62 &      0.132 &      0.286 \\
HolmbergII & (      143.899 ,       32.2721) & (      106.073 ,       49.5466) &     0.077 &           0.028 &       1.34 &     0.091 &      0.300 \\
M81DwarfA & (      143.536 ,       32.5857) & (      106.476 ,       49.8996) &     0.041 &          0.043 &  1.00 &      0.121 &      0.248 \\
DDO053 & (      148.991 ,       34.4963) & (      110.352 ,       45.7107) &     0.077 &           0.048 &   1.57 &      0.113 &      0.324 \\
NGC2798 & (      178.927 ,       43.7838) & (      127.750 ,       25.1694) &     0.051 &         0.040 &   1.04 &     0.086 &      0.166 \\
NGC2841 & (      166.458 ,       43.6210) & (      124.989 ,       33.8516) &      0.101 &          0.043 &  0.74 &     0.061 &      0.140 \\
NGC2976 & (      143.612 ,       40.4772) & (      118.782 ,       50.4361) &     0.076 &            0.100 &  2.59 &      0.594 &       1.951 \\
HolmbergI & (      140.502 ,       38.2525) & (      115.197 ,       52.8405) &     0.079 &           0.036 &       1.18 &      0.185 &      0.659 \\
NGC3049 & (      226.610 ,       44.4312) & (      146.934 ,      -2.98701) &     0.027 &             ---- &   2.20 &      0.223 &      0.504 \\
NGC3190 & (      211.891 ,       54.4481) & (      147.733 ,       10.7909) &     0.090 &             ---- &       1.61 &     0.094 &      0.195 \\
NGC3184 & (      178.537 ,       54.9707) & (      139.768 ,       28.3651) &      0.102 &           0.031 & 0.93 &     0.089 &      0.149 \\
NGC3198 & (      170.656 ,       54.2275) & (      138.036 ,       32.7413) &     0.050 &          0.039 &  0.52 &     0.074 &      0.157 \\
IC2574 & (      139.983 ,       43.1538) & (      123.351 ,       52.9795) &      0.156 &           0.036 &  1.38 &      0.205 &      0.509 \\
NGC3265 & (      200.457 ,       58.5603) & (      147.821 ,       18.3074) &     0.029 &          0.030 &  1.30 &     0.055 &      0.146 \\
MRK33 & (      156.610 ,       52.2144) & (      135.166 ,       41.0632) &     0.032 &          0.032 &      0.86 &      0.212 &      0.424 \\
NGC3351 & (      232.652 ,       56.1499) & (      157.328 ,       3.64920) &     0.074 &             ---- &       1.95 &      0.108 &      0.210 \\
NGC3521 & (      254.316 ,       52.8287) & (      166.849 ,      -5.14601) &      0.149 &              ---- &    2.81 &      0.328 &      0.718 \\
NGC3621 & (      280.580 ,       25.9208) & (      184.582 ,      -34.1228) &      0.180 &             0.111 & 4.12 &      0.444 &      0.900 \\
NGC3627 & (      240.246 ,       64.2844) & (      165.038 ,       8.27119) &     0.097 &             ---- &       1.70 &      0.179 &      0.248 \\
NGC3773 & (      249.132 ,       66.7215) & (      169.474 ,       8.73315) &     0.034 &             ---- &       1.84 &      0.102 &      0.125 \\
NGC3938 & (      153.689 ,       68.7273) & (      156.788 ,       39.2954) &     0.075 &          0.065 &    1.01 &      0.194 &      0.328 \\
NGC4125 & (      130.164 ,       50.9404) & (      139.741 ,       57.1897) &     0.095 &           0.037 &  0.81 &      0.174 &      0.285 \\
NGC4236 & (      127.310 ,       46.9854) & (      134.438 ,       60.6054) &      0.267 &           0.035 & 0.66 &     0.098 &      0.170 \\
NGC4254 & (      267.627 ,       75.3684) & (      177.739 ,       15.3315) &     0.075 &          0.054 &   2.37 &      0.161 &      0.370 \\
NGC4321 & (      267.852 ,       77.0715) & (      178.035 ,       17.0118) &      0.119 &           0.053 &   1.37 &      0.160 &      0.214 \\
NGC4450 & (      270.395 ,       78.8333) & (      178.801 ,       18.7003) &     0.079 &          0.032 &   1.36 &      0.119 &      0.338 \\
NGC4536 & (      291.454 ,       65.1324) & (      186.331 ,       5.66591) &      0.102 &             ---- &    1.64 &     0.090 &      0.239 \\
NGC4552 & (      285.274 ,       74.9059) & (      182.409 ,       14.8782) &      0.106 &             ---- &    2.28 &      0.210 &      0.688 \\
NGC4559 & (      193.095 ,       85.8409) & (      175.296 ,       29.2435) &      0.124 &           0.029 &  1.05 &     0.082 &      0.184 \\
NGC4569 & (      285.793 ,       75.9750) & (      182.385 ,       15.9551) &      0.101 &           0.060 &  2.65 &      0.174 &      0.470 \\
NGC4579 & (      287.803 ,       74.2939) & (      183.194 ,       14.3801) &     0.071 &            ---- &       2.37 &      0.126 &      0.311 \\
NGC4594 & (      297.305 ,       51.1460) & (      193.066 ,      -6.96614) &      0.151 &             ---- &     3.12 &      0.141 &      0.298 \\
NGC4625 & (      131.096 ,       75.1008) & (      168.734 ,       41.5922) &     0.047 &          0.052 &   0.88 &      0.121 &      0.239 \\
NGC4631 & (      145.167 ,       83.5710) & (      174.202 ,       33.9219) &      0.171 &          0.054 &  1.02 &      0.182 &      0.248 \\
NGC4725 & (      271.647 ,       88.6791) & (      179.905 ,       28.4957) &     0.079 &         0.037 &   0.70 &      0.102 &      0.161 \\
NGC4736 & (      124.738 ,       75.5220) & (      170.796 ,       42.2419) &      0.187 &          0.047 &  0.67 &      0.192 &      0.293 \\
DDO154 & (      110.504 ,       89.5114) & (      179.918 ,       30.2879) &     0.042 &           0.043 &   0.57 &     0.085 &      0.137 \\
NGC4826 & (      311.183 ,       85.0014) & (      183.181 ,       25.6681) &      0.125 &          0.059 &  1.89 &      0.182 &      0.239 \\
DDO165 & (      121.301 ,       49.1180) & (      142.645 ,       62.9864) &     0.058 &           0.051 &   1.14 &      0.171 &      0.301 \\
NGC5033 & (      101.051 ,       79.6584) & (      178.930 ,       40.1141) &      0.128 &          0.033 &  0.54 &     0.096 &      0.149 \\
NGC5055 & (      107.903 ,       73.9458) & (      175.505 ,       45.4687) &      0.163 &          0.052 &  0.75 &      0.178 &      0.267 \\
NGC5194 & (      106.114 ,       68.6881) & (      174.458 ,       50.7106) &      0.233 &          0.087 &  1.23 &      0.364 &      0.712 \\
Tololo89 & (      318.732 ,       27.7794) & (      219.272 ,      -19.6125) &     0.051 &             0.186 &       5.26 &      0.544 &      0.978 \\
NGC5408 & (      316.804 ,       20.0386) & (      223.038 ,      -26.7581) &     0.056 &             0.163 &  5.55 &      0.761 &       1.702 \\
NGC5474 & (      101.503 ,       59.8880) & (      174.929 ,       59.7255) &     0.061 &           0.027 & 0.48 &     0.086 &      0.185 \\
NGC5713 & (      350.094 ,       52.5981) & (      217.013 ,       14.3062) &     0.052 &             ---- &       2.08 &     0.070 &      0.187 \\
NGC5866 & (      92.3464 ,       52.7620) & (      186.246 ,       66.8587) &      0.105 &           0.029 & 0.62 &     0.097 &      0.210 \\
IC4710 & (      327.879 ,      -22.1180) & (      273.178 ,      -43.4420) &     0.034 &             0.119 &   5.69 &      0.426 &       1.007 \\
NGC6822 & (      24.6969 ,      -17.9857) & (      294.785 ,       6.07295) &      0.185 &            ---- &       7.77 &      0.745 &       1.705 \\
NGC6946 & (      95.3945 ,       12.0399) & (      356.855 ,       72.2046) &     0.086 &      0.125 &       9.57 &      0.722 &       1.948 \\
NGC7331 & (      93.0851 ,      -20.9465) & (      356.050 ,       39.1427) &      0.126 &           0.121 &   3.91 &      0.429 &      0.761 \\
NGC7552 & (      347.564 ,      -64.7400) & (      330.605 ,      -34.6836) &     0.046 &          0.060 &  0.81 &      0.128 &      0.173 \\
NGC7793 & (      5.28269 ,      -76.5898) & (      344.625 ,      -29.1746) &      0.119 &           0.046 & 0.95 &      0.126 &      0.184 \\
\enddata
\end{deluxetable}

\begin{deluxetable}{c c c}
\tabletypesize{\scriptsize}
\tablecaption{Infrared colors in the observations, i.e. the 160/100 and 60/100 brightness ratios obtained from a fit of the correlations in each field\label{tab2}}
\tablewidth{0pt}
\tablehead{
\colhead{field} & \colhead{160/100 color} &  \colhead{60/100 color} }
\startdata
NGC0337 &       $      2.3\pm     0.2$ &       --- \\
NGC0628 &       $      2.6\pm     0.1$ &       --- \\
NGC0855 &       $      2.7\pm      0.5$ &       --- \\
NGC0925 &       $      2.4\pm      0.3$ &      $     0.12\pm     0.03$ \\
NGC2976 &       $      3.0\pm     0.1$ &      $     0.15\pm    0.01$ \\
NGC3521 &       $      1.81\pm     0.07$ &      --- \\
NGC3621 &       $      1.86\pm     0.04$ &      $     0.225\pm    0.005$ \\
NGC4569 &       $      2.0\pm      0.2$ &       $     0.31\pm     0.03$ \\
NGC4594 &       $      2.3\pm      0.1$ &       --- \\
TOL89 &       $      1.7\pm     0.1$ &       $     0.32\pm    0.02$ \\
NGC5408 &       $      2.22\pm     0.04$ &      $     0.19\pm    0.01$ \\
IC4710 &       $      2.0\pm      0.1$ &       $     0.31\pm     0.01$ \\
NGC6822 &       $      2.23\pm     0.04$ &      --- \\
NGC6946 &       $      2.5\pm     0.1$ &      $     0.16\pm    0.006$ \\
NGC7331 &       $      1.75\pm     0.07$ &      $     0.238\pm    0.009$ \\
\enddata
\end{deluxetable}

\end{document}